\documentclass[10pt, aps,prc,twocolumn,superscriptaddress,preprintnumbers,
amsmath, 
floatfix,
longbibliography,
nofootinbib
]{revtex4-1}
\usepackage[T1]{fontenc}
\usepackage[utf8x]{inputenc} 
\usepackage{adjustbox}          
\usepackage[caption=false]{subfig}
\usepackage{url}
\usepackage{color}
\usepackage{float}
\usepackage[pdftex,colorlinks=true, linkcolor = blue, citecolor=blue,urlcolor=blue, bookmarksnumbered=true, bookmarksopen=true]{hyperref}
\usepackage{longtable}
\usepackage{amsfonts}
\usepackage{dsfont}
\usepackage{wrapfig,bm} 
\usepackage[normalem]{ulem}
\usepackage{MnSymbol}

\newcommand{\beq}{\begin{equation}}
\newcommand{\eeq}{\end{equation}}
\newcommand{\bea}{\begin{eqnarray}}
\newcommand{\eea}{\end{eqnarray}}

\begin{document}

\title{Non-Markovian character and irreversibility of real-time quantum many-body dynamics}


\author{Aurel Bulgac}%
 \author{Matthew Kafker}   
 \affiliation{Department of Physics,%
  University of Washington, Seattle, WA 98195--1560, USA}

 \author{Ibrahim Abdurrahman}
 \author{Ionel Stetcu}
 \affiliation{ Theoretical Division, Los Alamos National Laboratory, Los Alamos, NM 87545, USA}
 
\date{\today}

\begin{abstract}

The presence of pairing correlations within the time-dependent density functional theory (TDDFT) extension to superfluid systems, is tantamount to the presence of a quantum collision integral in the evolution equations, which leads to an obviously non-Markovian behavior of the single-particle occupation probabilities, unexpected in a traditional quantum extension of kinetic equations. The quantum generalization of the Boltzmann equation, based on a collision integral in terms of phase-space occupation probabilities, is the most used approach to describe nuclear dynamics and which by construction has a Markovian character.
By contrast, the extension of TDDFT to superfluid systems has similarities with the Baym and Kadanoff kinetic formalism, which however is formulated with much more complicated evolution equations with long-time memory terms and non-local interactions. The irreversibility of quantum dynamics is properly characterized using the canonical wave functions/natural orbitals, and the associated canonical occupation probabilities, which provide the smallest possible representation of any fermionic many-body wave function. In this basis, one can evaluate the orbital entanglement entropy, which is an excellent measure of the non-equilibrium dynamics of an isolated system. To explore the phenomena of memory effects and irreversibility, we investigate the use of canonical wave functions/natural orbitals in nuclear many-body calculations, assessing their utility for static calculations, dynamics, and symmetry restoration. As the number of single-particle states is generally quite large, it is highly desirable to work in the canonical basis whenever possible, preferably with a cutoff. We show that truncating the number of canonical wave functions can be a valid approach in the case of static calculations, but that such a truncation is not valid for time-dependent calculations, as it leads to the violation of continuity equation, energy conservation, and other observables, as well as an inaccurate representation of the dynamics. Indeed, in order to describe the dynamics of a fissioning system within a TDDFT framework all the canonical states must be included in the calculation. Finally, we demonstrate that the canonical representation provides a very efficient basis for performing symmetry projections.

\end{abstract} 

\preprint{NT@UW-23-17}
\preprint{LA-UR-23-33343}

\maketitle


\section{Introduction} \label{sec:I}

We address the question of whether nuclear dynamics has a Markovian character or not, and the related aspect, whether the dynamics of a typically excited isolated nucleus is irreversible and how to describe its irreversibility. As we will show, the main technical tool needed to address these issues is the set of canonical wave functions or natural orbitals.

Canonical wave functions were introduced in 1957 by Bardeen, Cooper, and Schrieffer to describe superconductors \cite{Bardeen:1957}.  A year earlier in 1956 L\"owdin introduced a very similar set of single-particle wave functions, which he called natural orbitals ~\cite{Lowdin:1956a,Lowdin:1956,Coleman:1963,Coleman:1963a,Davidson:1972}. It was proven~\cite{Lowdin:1956,Lowdin:1956a,Coleman:1963}  that the canonical wave functions/natural orbitals can serve as the most economical way to represent any many-body wave function as a sum over Slater determinants, each one of which obviously describes a system of $N$ independent fermions. The sum of many such Slater determinants however describes in general a strongly interacting system of particles. After determining the set of canonical wave functions/natural orbitals one can extract a unique and well-defined set of single-particle occupation probabilities, known as the canonical occupation probabilities, which allows one to evaluate the orbital entanglement entropy of the many-body wave function, which serves as a basis-independent characterization of its complexity. While such a set of single-particle wave functions is likely the best way to establish the complexity of a many-body wave function of an interacting system, the use of such set become problematic for time-dependent or non-equilibrium situations. The reason is quite obvious: due to particle-particle interactions the single-particle occupation probabilities are time-dependent, and so are also the canonical wave functions/natural orbitals, and so the appealing first thought that one can use the canonical wave functions defined at the initial time in time-dependent problems proves false right away, even after one time step.
 
\textcite{Bardeen:1957} (BCS) were the first to introduce a highly correlated many-body wave function to 
describe fermionic superfluids, immediately followed by similar suggestions due to
\textcite{Bogoliubov:1958} and \textcite{Valatin:1958}.  The BCS wave function has the form
\begin{align}
&|\Phi\rangle =  \prod_k(u_k+v_ka^\dagger_ka^\dagger_{\overline k})|0 \rangle, \quad |u_k|^2+|v_k|^2=1, \label{eq:BCS}\\
&\phi_k(\xi)= \langle \xi |a^\dagger_k|0\rangle, \, \phi_{\overline k}(\xi)= \langle \xi |a^\dagger_{\overline k}|0\rangle,
\end{align} 
where $a^\dagger_k$ $a^\dagger_{\overline k}$ are single-particle creation operators for time-reversed  states, $|0\rangle$ 
is the vacuum state, and $\phi_k(\xi), \, \phi_{\overline k}(\xi)$ are the corresponding single particle wave 
functions depending on the spatial and spin (and isospin for nuclear systems)  coordinate $\xi = ({\bm r}, \sigma,\tau)$, known as the canonical wave functions.

In general,  one follows Bogoliubov and now introduces a more general type of many-body wave function for a many-body system 
with pairing correlations, using creation and annihilation quasi-particle operators and the corresponding many-body wave function 
\begin{align} 
\!\!\!\!\!\! &\alpha_k^\dagger  =  
\sumint_\xi \left [ {\textrm u}_k(\xi) \psi^\dagger (\xi) + {\textrm v}_k(\xi) \psi (\xi)\right ], \label{eq:a0}\\
\!\!\!\!\!\! &\alpha_k= 
\sumint_\xi\left [ {\textrm v}_k^*(\xi) \psi^\dagger (\xi ) + {\textrm u}_k^*(\xi) \psi (\xi)\right ], \label{eq:b0}\\
\!\!\!\!\!\! & |\Phi\rangle = {\cal N}\prod_k \alpha_k|0\rangle,
\end{align}
where ${\cal N}$ is an appropriate normalization constant and with the reverse relations  
\begin{align} 
&\psi^\dagger (\xi) = \sumint_k \left [ {\textrm u}^*_k(\xi)  \alpha^\dagger _k  
                                            + {\textrm v}_k(\xi)\alpha_k \right ], \label{eq:p1}\\
&\psi(\xi) =                \sumint_k \left [ {\textrm v}^*_k(\xi)\alpha^\dagger_k
                                            + {\textrm u}_k(\xi)\alpha_k \right ], \label{eq:p2} 
\end{align}
where $\psi^\dagger (\xi)$ and $ \psi (\xi)$ are the field operators 
for the creation and annihilation of a particle with coordinate $\xi=({\bm r},\sigma,\tau)$, 
$(u_k(\xi),v_k(\xi))^{T}$ are the quasi-particle wave functions,
and the integral implies also a summation over discrete variables when appropriate.
The quasi-particle wave functions ${\textrm u}_k(\xi), {\textrm v}_k(\xi)$ are determined 
by solving the Hartree-Fock-Bogoliubov (HFB) equations and are eigenstates 
of the corresponding HFB quasi-particle Hamiltonian. 

The BCS approximate many-body wave function is an excellent candidate for an electronic superconductor, when the pairing correlations 
are limited to a very narrow energy interval around the Fermi level, and the single-particle wave functions have a negligible energy dependence. 
For nuclei, neutron and proton matter in neutron stars, and cold atoms however, when the pairing interactions are strong and the mixing occurs 
among states rather well separated in single-particle energy, the assumption that the $u_k(\xi)$ and $v_k(\xi)$  components of the quasi-particle 
wave functions have similar spatial dependence is not valid anymore.
In the Hartree-Fock-Bogoliubov  approximation, the quasi-particle components $u_k(\xi)$ and $v_k(\xi)$ have very different spatial behavior. 
In particular while the $v_k(\xi)$ components in the case of nuclei  or isolated finite systems are always square integrable~\cite{Bulgac:1980}, 
the $u_k(\xi)$ components most of the time are continuum type of wave functions, which are not square integrable. 

In Section~\ref{sec:II} we will describe how canonical wave functions/natural orbitals are defined, subsection~\ref{ssec:IIA},  and relevant aspects of the Bloch-Messiah decomposition,  subsection~\ref{ssec:IIB}. In Section~\ref{sec:III} we will discuss generalized Bogoliubov quasiparticles, which become relevant in reactions, when both partners are superfluid, as in the case of nuclear fission or collision between heavy-ions. In Section~\ref{sec:IV} we will discuss several particular aspects which are relevant in the subsequent analysis of the time-dependent equations for fermionic superfluids. In Section~\ref{sec:V} we describe some new aspects of the particle number projection for fermionic superfluids, which will be illustrated in the case of nuclear fission in the following section. In subsection~\ref{ssec:VIA} we discuss under what circumstances the use of a reduced set of canonical wave functions is beneficial.  In subsection~\ref{ssec:VIB}, we discuss the non-Markovian character of the fission dynamics. In subsection~\ref{ssec:VIC}, we discuss the relevance of particle number projection and also the use of the reduced set of canonical wave functions in dynamic simulations. And in subsection~\ref{ssec:VID} we will illustrate the irreversible fission dynamics and its  characterization by the means of the orbital entanglement entropy. In Section~\ref{sec:VII} we will summarize our main findings. 

\section{Canonical wave functions} \label{sec:II}

The set of quasi-particle wave functions ${\textrm u}_k(\xi), {\textrm v}_k(\xi)$ is twice the size of the set of canonical wave functions  $\phi_k(\xi), \, \phi_{\overline k}(\xi)$. These 
two sets of quasi-particle wave functions however are related and one can derive one set from 
the other and vice versa, see Ref.~\cite{Bulgac:2023} and the text below. Practice shows that using a significantly reduced set of functions
$\phi_k(\xi), \, \phi_{\overline k}(\xi)$, with occupation numbers $n_k=|v_k|^2$ is often sufficient to represent the many-body wave function $|\Phi\rangle$ with
sufficient accuracy ~\cite{Bulgac:2023}. The quasi-particle 
wave functions  ${\textrm u}_k(\xi), {\textrm v}_k(\xi)$ are very useful to describe low-energy excitations 
of the many-body system and their corresponding eigenvalues $E_k\geq 0$ play a similar role as the 
particle eigenstates in a normal system. 

The canonical wave functions can be determined after diagonalizing the overlap Hermitian matrix 
\begin{align}
{\cal O}_{kl}= \langle {\text v}_k|{\text v}_l\rangle, \label{eq:overlap}
\end{align}
and the resulting canonical $\tilde{v}_k$ components, defined below,  satisfy the relations
\begin{align} 
\langle\tilde{{\textrm v}}_k|\tilde{{\textrm v}}_l\rangle = n_k\delta_{kl,} \label{eq:noc}
\end{align}
where $n_k$ are the canonical occupation probabilities~\cite{Bulgac:2023}.
It follows that the overlap matrix of the $\tilde{\textrm u}_k$-components is also diagonal
\begin{align} 
\langle\tilde{{\textrm u}}_k|\tilde{{\textrm u}}_l\rangle = (1-n_k)\delta_{kl},
\end{align}
and the average particle number is given by 
\begin{align} 
N = \sum_k n_k = \sum_k \langle\tilde{{\textrm v}}_k|\tilde{{\textrm v}}_k\rangle =\sum_k \langle{{\textrm v}}_k|{{\textrm v}}_k\rangle.
\end{align}
One should note that the occupation probabilities $n_k=\langle\tilde{{\textrm v}}_k|\tilde{{\textrm v}}_k\rangle$ are different from 
$\langle{\textrm v}_k|{\textrm v}_k\rangle$, these latter ones not being gauge invariant. As a rule, the eigenvalues $n_k$ are double degenerate for even-even nuclei 
and, depending on the formulation, the eigenvalue spectrum is either discrete for systems in a finite box or a mixture of discrete 
and continuous spectrum in an infinite box.
The number of ${\textrm v}_k$-components  for either the proton or neutron subsystems 
is $2\Omega =2N_xN_yN_z$ for neutrons and protons respectively in a finite box, 
where $N_{x,y,z}$ are the number of lattice points in the corresponding cartesian direction.
 
It is useful to introduce the unitary transformation, and correspondingly the set of eigenvectors, which diagonalizes $O_{kl}$
\begin{align}
&\sum_lO_{kl}{\cal U}_{lm} = {\cal U}_{km}n_m, \quad \sum_n{\cal U}^*_{km}{\cal U}_{kn} = \delta_{mn},\\
&O_{kl} = \sum_m{\cal U}_{km}n_m{\cal U}^*_{lm},\\
&{\textrm v}_k(\xi)=\sum_m{\cal U}_{km}^*\tilde{{\text v}}_m(\xi),  \quad \tilde{\textrm v}_n(\xi)=\sum_l{\cal U}_{ln}{\textrm v}_l(\xi)\label{eq:v}\\
&{\textrm u}_k(\xi)=\sum_m{\cal U}_{km}^*\tilde{{\textrm u}}_m(\xi), \quad \tilde{\textrm u}_n(\xi)=\sum_l{\cal U}_{ln}{\textrm u}_l(\xi).\label{eq:uv0}
\end{align}
The columns of the matrix ${\cal U}_{mn}$ are the eigenvectors of the overlap matrix, and they are used to construct the canonical wave functions $\tilde{\textrm v}_n(\xi)$ and $\tilde{\textrm u}_n(\xi)$. As before, the spatial, spin, and isospin coordinates are labeled by $\xi=({\bm r},\sigma,\tau)$.

Recently \textcite{Chen:2022} implemented a solution of the HFB self-consistent equations directly in the canonical basis set of wave functions. Since the canonical 
wave functions are not the same as the quasi-particle wave functions, solving explicitly the HFB equation within such a basis set requires the introduction 
of a large number of Lagrange multipliers, which have to be evaluated at each iteration.  One must re-diagonalize 
the canonical wave functions at each iteration, which makes a very cumbersome numerical implementation for large sets of canonical wave functions, 
which are needed in practice, as we show below. 
By contrast, solving the HFB equations self-consistently using standard diagonalization methods 
is a rather simple procedure, which is widely used in practice with rather large basis sets. After completing the iterative procedure, the determination 
of the canonical basis set requires only one diagonalization of the  overlap matrix defined in Eq.~\eqref{eq:overlap}, 
which is only half the size of the HFB matrix, followed by a unitary transformation from 
the original quasi-particle wave functions to the new canonical quasi-particle wave functions, see Eqs.~(\ref{eq:v}, \ref{eq:uv0}). 

\subsection{Relations between the canonical wave functions $\phi_{l,\pm \tau}(\xi)$  
and the quasi-particle wave functions $\tilde{\textrm u}_{l,\pm\tau}(\xi), \tilde{\textrm v}_{l,\pm\tau}(\xi)$} \label{ssec:IIA}

The canonical occupation probabilities and wave functions for an even system are defined as
\begin{align}
& n(\xi,\xi') = \sum_k {\text v}^*_k(\xi){\text v}_k(\xi'),\\
&\sumint_{\xi'}d\xi' n({\xi},\xi')\phi_{l,\tau}(\xi') = {\text n}_l\phi_{l,\tau}(\xi), \label{eq:canwf}\\
&\sum_{l,\tau} \phi_{l,\tau}(\xi) \phi^*_{l,\tau}(\xi')=\delta_{\xi,\xi'}, \\
& \langle  \phi_{l,\tau} | \phi_{l',\tau'} \rangle = \delta_{l,l'} \delta_{\tau,\tau'},
\end{align}
where $\tau = \pm$, $l=1,...,\Omega$,  and  ${\text n}_l$ are double degenerate. 
This definition of the canonical wave functions has an ambiguity, as their phases are undefined. While the overall phases of the $\tilde{\text v}_k$ components 
are irrelevant for the definition of the normal number densities, the relative phases of the $\tilde{\text u}_k$ and $\tilde{\text v}_k$ components are 
crucial for the correct reproduction of the anomalous density.  The set of wave functions $\phi_{l,\tau}(\xi)$ 
can be introduced for any many-body system and as such were introduced first by L\"owdin in 1956 
and called natural orbitals~\cite{Lowdin:1956a,Lowdin:1956,Davidson:1972,Coleman:1963,Coleman:1963a} and it can be shown that they represent 
the most economical way to represent accurately a many-body wave function.

Using canonical WFs one can show that the many-body wave function $|\Phi\rangle$  has the structure (up to an overall irrelevant phase)
\begin{align}
|\Phi\rangle = \prod_{l=1}^{\Omega}(\sqrt{1-{\text n}_l} + \sqrt{{\text n}_l}a^\dagger_{l,+}a^\dagger_{l,-})|0\rangle \label{eq:Phi}
\end{align}
where $a^\dagger_{l,\tau}$ are creation operators for the canonical wave functions $\phi_{l,\tau}(\xi)=\langle\xi|a^\dagger_{l,\tau}|0\rangle$ defined in Eq.~\eqref{eq:canwf},
and  that $|\Phi\rangle$ is a quasi-particle vacuum for the  canonical quasi-particle operators $\alpha_{l,\tau}|\Phi\rangle = 0$, 
\begin{align}
 \begin{pmatrix} \alpha^\dagger_{l,+} \\ \alpha^\dagger_{l,-} \\ \alpha_{l,+} \\ \alpha_{l,-} \end{pmatrix}=
\begin{pmatrix}  &0 & u_p & v_p &  0 \\ &u_p & 0 & 0 & -v_p \\ & v_p & 0 &0 & u_p\\ &0 & -v_p& u_p &0 \end{pmatrix} 
\begin{pmatrix} a^\dagger_{l,+}\\ a^\dagger_{l,-}\\ a_{l,+}\\ a_{l,-}\end{pmatrix}, \label{eq:can0} \\
\begin{pmatrix} a^\dagger_{l,+}\\ a^\dagger_{l,-}\\ a_{l,+}\\ a_{l,-}\end{pmatrix}=  
\begin{pmatrix}  &0 & u_p & v_p &  0 \\ &u_p & 0 & 0 & -v_p \\ & v_p & 0 &0 & u_p\\ &0 & -v_p& u_p &0 \end{pmatrix} 
 \begin{pmatrix} \alpha^\dagger_{l,+} \\ \alpha^\dagger_{l,-} \\ \alpha_{l,+} \\ \alpha_{l,-} \end{pmatrix},\label{eq:can1}
\end{align}
where $p=(l,+)$ and  $\overline{p} =(l,-)$ and $u_p = \sqrt{1-n_l}>0$ and $v_p = \sqrt{n_l}>0$ 
are assumed to be non-negative~\cite{Ring:2004,Bloch:1962,Zumino:1962,Bulgac:2021c}.
One can now introduce the corresponding quasi-particle wave functions
\begin{align}
&\tilde{\alpha}^\dagger_{l,\tau} = \sumint_\xi[\tilde{\text u}_{l,\tau}(\xi)\psi^\dagger(\xi)+
                                                             \tilde{\text v}_{l,\tau}(\xi)\psi(\xi)],\\
&\tilde{\alpha}_{l,\tau} =               \sumint_\xi[\tilde{\text v}^*_{l,\tau}(\xi)\psi^\dagger(\xi)+
                                                             \tilde{\text u}^*_{l,\tau}(\xi)\psi(\xi)], \\                                                             
&\tilde{\text u}_{l,\tau}(\xi) =\sqrt{1-{\text n}_l}\phi_{l,-\tau}(\xi), \label{eq:uqp}\\
&\tilde{\text v}_{l,\tau}(\xi) =\tau\sqrt{{\text n}_l}, \phi^*_{l,\tau}(\xi), \label{eq:vqp} 
\end{align}
or in matrix form
\begin{align}
\begin{pmatrix}\tilde{\alpha}^\dagger_{l,\tau} \\ \tilde{\alpha_{l,\tau}}\end{pmatrix} = \sumint_\xi
\begin{pmatrix} \tilde{\text u}_{l,\tau}(\xi)  \quad \tilde{\text v}_{l,\tau}(\xi)  \\
                          \tilde{\text v}^*_{l,\tau}(\xi) \quad \tilde{\text u}^*_{l,\tau}(\xi)
\end{pmatrix}
\begin{pmatrix} \psi^\dagger(\xi) \\ \psi(\xi) \end{pmatrix}. 
\end{align}

The time-reversal  symmetry between the two $\phi_{l,\tau}(\xi)$ of the doublet is not a necessary condition  in order 
to uniquely evaluate the anomalous density, as one can show that a unitary transformation between 
the two time-reversed canonical wave functions  $\phi_{l,\pm\tau}(\xi)$ does not change the Cooper pair wave function 
or the anomalous density 
\begin{align}
&\Psi_2(\xi,\zeta)
= {\cal N}\sum_l \left [  \tilde{\text v}^*_{l,\tau}(\xi)     \tilde{\text u}_{l,\tau}(\zeta) 
                                                    -\tilde{\text v}^*_{l,\tau}(\zeta)\tilde{\text u}_{l,\tau}(\xi) \right ],\label{eq:Cooper}
\end{align} 
as the quantity in square brackets is a $2\times2$ Slater determinant, and
where ${\cal N}$ is the normalization constant. This wave function is invariant with respect to an arbitrary unitary transformation between 
$\phi_{l,\pm \tau}(\xi)$ for fixed $l$, and therefore $\phi_{l,\pm \tau}(\xi)$ do not need to be related by time-reversal symmetry.

Since the anomalous density 
$\kappa(\xi,\xi')=-\kappa(\xi',\xi)$ is by definition antisymmetric,  the representation of the many-body wave function 
in the canonical basis is well defined only if the two canonical wave functions $\phi_{l,\pm\tau}(\xi)$, 
and correspondingly the wave functions $ \tilde{\text u}_{l,\tau}(\xi),  \tilde{\text v}_{l,\tau}(\xi)$, have a well defined relative phase
\begin{align}
&\tilde{\text u}_{l,-\tau}(\xi) = \tau\sqrt{ \frac{1-{\text n}_l}{ {\text n}_l} }\tilde{\text v}^*_{l,\tau}(\xi),\\
&\tilde{\text u}_{l,\tau}(\xi) = -\tau\sqrt{\frac{ 1-{\text n}_l}{ {\text n}_l} }\tilde{\text v}^*_{l,-\tau}(\xi),
\end{align}
and if the WFs $\tilde{\text u}_k(\xi)$ and $\tilde{\text v}_k(\xi)$ are defined by Eqs.~(\ref{eq:v}, \ref{eq:uv0}). One can check that these 
relations ensure the antisymmetry of 
\begin{align} 
\kappa(\xi,\zeta) &= \sum_{l,\tau}  \tilde{\text v}^*_{l,\tau}(\xi)\tilde{\text u}_{l,\tau}(\xi) \nonumber \\
&= \sum_{l,\tau} \tau \sqrt{ \text{n}_l(1-\text{n}_l) } \phi_{l,\tau}(\xi) \phi_{l,-\tau}(\zeta).
\end{align}
The issue with using the functions $\phi_{l,\tau}(\xi), \phi_{l,-\tau}(\zeta) $ as independent eigenfunctions is their relative phase ambiguity. 
The canonical $\phi_{l}(\xi) = \phi_{l,+}(\xi)$ and $\phi_{\overline{l}}(\xi) = \phi_{l,-}(\xi)$ with correct relative phases are
 \begin{equation}
 \phi_{l}(\xi)= \frac{\tilde{\text v}^*_{l,+}(\xi)}{\sqrt{{\text n}_l}},\quad \phi_{\overline{l}}(\xi) = \frac{\tilde{\text u}_{l,-}(\xi) }{\sqrt{1-{\text n}_l}}.
 \end{equation}
 Specifying these phases correctly is necessary to obtain correct answers when computing overlaps between different HFB vacua using Pfaffians~\cite{Robledo:2009,Bertsch:2012,Carlsson:2021}.  Furthermore, the quasi-particle wave functions $\tilde{\text u}_{l,\tau}(\xi)$ and 
 $\tilde{\text v}_{l,\tau}(\xi)$ are useful for evaluating density matrices between different HFB vacua.

\subsection{Bloch and Messiah decomposition}\label{ssec:IIB}

\textcite{Bloch:1962,Ring:2004} used the following relations between the quasi-particle and field operators 
(note that these authors used a reverse order of creation and annihilation operators) 
\begin{align}
&\left ( \begin{array}{c} \beta^\dagger \\ \beta \end{array}\right ) = 
{\cal W}^\dagger  \left ( \begin{array}{c} \psi^\dagger \\ \psi \end{array}\right )
 =\left ( \begin{array}{cc}  {U}^T&{V}^T \\ V^\dagger&U^\dagger \end{array}  \right )\left ( \begin{array}{c} \psi^\dagger \\ \psi \end{array}\right ),\\
&{\cal W}= \left( \begin{array}{cc} U^*&V\\V^*&U\end{array} \right )
= \left( \begin{array}{cc} D^*& 0 \\ 0 &D\end{array} \right )
\left( \begin{array}{cc} \overline{U}&\overline{V}\\ \overline{V}&\overline{U}\end{array} \right )
\left( \begin{array}{cc} C^*&0\\0&C\end{array} \right ),\\
&{\cal W}^\dagger 
= \left( \begin{array}{cc} {C}^T& 0 \\ 0 &C^\dagger\end{array} \right )
\left( \begin{array}{cc} \overline{U}&\overline{V}\\ \overline{V}&\overline{U}\end{array} \right )
\left( \begin{array}{cc} {D}^T&0\\0&D^\dagger\end{array} \right ),
\end{align}
where we have dropped the quasi-particle labels and particle coordinates $\xi$, 
$\overline{U}$ is a $2\times2$ real block-diagonal matrix and $\overline{V}$ is a $2\times2$ real block skew-symmetric matrix if $0<n_k<1$.
Otherwise, if  $n_k=0$ or 1 the corresponding $2\times2$ matrices are real-diagonal.   Comparing these relations with Eqs.~(\ref{eq:v}, \ref{eq:uv0}) 
it is easy to see that the matrix ${C}^{-1}= {\cal U}^T$. We use ${M}^T$ for  a transpose of a matrix ${ M}$. 

\section{Generalized Bogoliubov quasi-particles}\label{sec:III}

We will describe here a generalization of the Bogoliubov quasi-particle creation and annihilation operators, which can be useful in several applications, which we describe below.
Let us separate the space into two regions defined by the Heaviside functions
\begin{align}
&\Theta_L(\xi)+\Theta_R(\xi)=1,\quad \Theta_L (\xi)\Theta_R(\xi)=0,\\
&\Theta_R(\xi) = 
\left \{
\begin{matrix}
1 \quad z\le 0,\\
0 \quad  z>0.
\end{matrix} 
\right . 
\end{align}
and introduce also the field operators 
\begin{align}
\psi^\dagger _{L,R}(\xi) = \psi^\dagger(\xi)\Theta_{L,R}(\xi).
\end{align}
The separation of the space into two parts can be perform in any manner, e.g. a Swiss cheese type, with hole belonging to one part and the filled part to the other.

One can then define the generalized Bogoliubov quasi-particles
\begin{align}
& \begin{pmatrix} \alpha_{L,k}^\dagger \\ \alpha_{R,k}^\dagger\\ \alpha_{L,k} \\ \alpha_{R,k} \end{pmatrix} \label{eq:gcan}\\
&=\sumint_\xi \begin{pmatrix}  & u_{L,k}(\xi)& 0                                & v_{L,k}(\xi) &  0 \\  
                                             & 0                             &u_{R,k}(\xi)   & 0               & v_{L,k}(\xi) \\ 
                                             & v_{L,l}^*(\xi)  & 0                               & u_{L,k}^*(\xi) & 0\\ 
                                             & 0                              & v_{R,k}^*(\xi)  &                  &u_{R,k}^*(\xi) \ \end{pmatrix} 
\begin{pmatrix} \psi^\dagger_L(\xi) \\  \psi^\dagger_R(\xi) \\ \psi_L(\xi)\\ \psi_R(\xi)\end{pmatrix},\nonumber
\end{align}
and one can then define the many-body wave function
\begin{align}
|\Phi \rangle = {\cal N} \prod_k \alpha_{L,k}\alpha_{R,k}|0\rangle,
\end{align}
which will describe two uncorrelated superfluid fermion systems in two different parts of the space.
By defining a new set of Bogoliubov quasi-particles through a unitary transformation
\begin{align}
\begin{pmatrix} 
\alpha^\dagger_{k,+}\\
\alpha^\dagger _{k,-}\\
\end{pmatrix} 
= \sum_l 
\begin{pmatrix}
& U_{k,l} & V_{k,l} \\
& V^*_{k,l}&U^*_{k,l}
\end{pmatrix}
\begin{pmatrix} 
\alpha^\dagger_{L,l}\\
\alpha^\dagger_{R,l}
\end{pmatrix}
\end{align}
and a similar transformation for the corresponding annihilation operators one recovers 
the original definition of the Bogoliubov creation and annihilation operators and the usual definition of the quasi-particle vacuum
\begin{align}
|\Phi\rangle ={\cal N} \prod_k \alpha_{k,+}\alpha_{k,-}|0\rangle.
\end{align} 
These new type of Bogoliubov quasi-particles are useful when studying the importance of the relative phase 
between two condensates, prepared either independently or in 
a controlled manner as discussed in Refs.~\cite{Magierski:2017,Magierski:2022,Bulgac:2017}.

The interaction of two superfluids, which at some point in time are spatially separated, is a quantum problem likely even 
more mysterious than quantum entanglement. To appreciate how unusual this problem is one has to invoke the description of
either Bose-Einstein condensates or fermionic superfluid systems. In both cases one introduces the Bogoliubov quasi-particles, 
see Eqs.~(\ref{eq:a0}, \ref{eq:b0}), in which one has the ${\text u}_k(\xi)$- and the  ${\text v}_k(\xi)$-components of 
the quasi-particle wave functions.   The ${\text v}_k(\xi)$-component is a wave function of a spin 1/2 particle, for which according 
to Max Born's Quantum Mechanics ``dogma''~\cite{Born:1926a,Born:1926b,Heilbron:2023},
the quantity   $|{\text v}_k({\bm r},\sigma,\tau)|^2 d^3r$ is interpreted as 
the probability to find a fermion with spin $\sigma$ and  isospin $\tau$, in the 3D spatial volume $d^3r$, and which in 
principle can be measured, in a similar manner as the spin $\sigma$ or the isospin $\tau$ can be measured. On the other hand 
it is totally unclear what the ${\text u}_k(\xi)$-component describes, apart from the fact that $\sumint_\xi |{\text u}_k({\bm r},\sigma,\tau)|^2 d^3r$
is the probability for the specific single-particle quantum state to be unoccupied, basically a ``ghost particle.''  In spite of this 
lack of interpretation of the components of the quasi-particle wave functions, for many decades now theorists happily use the 
Hartree-Fock-Bogoliubov (HFB) approximation without ever wondering what is the meaning of  the ${\text u}_k({\bm r},\sigma,\tau)$-component of the quasi-particle wave function. 
Moreover, the pairing potential explicitly depends on these ${\text u}$-components and 
it extends in space, much further than the matter distribution of the many-body fermion wave function~\cite{Bulgac:1980}. Therefore, the ground state properties 
evaluated in the HFB approximation depend in a critical manner on the ``mysterious'' properties of the ${\text u}$-components of the 
quasi-particle wave functions, which ``describe'' the probability to find a ``ghost particle,'' with corresponding wave functions  
which have coordinates and even a time-dependence, with a finite probability to ``find'' it in  little volume in space $d^3r$. 

The list of questions arising in the treatment of superfluids, which have not been asked yet, is even longer.  
Assume that a fermionic superfluid, and even a bosonic one as well~\cite{Anderson:1986,Bulgac:2017}, approaches another normal 
system, and it is a relatively ``safe'' distance such that the matter distributions of the two systems have a negligible overlap 
and that the particle-particle interaction is short-ranged. The pairing field of 
the superfluid system however, since it extends well beyond its own matter distribution, creates an ``external'' pairing field in the normal system and 
as a result pairing correlations are induced in the normal system. In condensed matter systems, similar situations are well-known at the 
interface of a superfluid and an insulator, but in that case the ``border region'' between the superfluid and the insulator is of the order 
of the atomic distances and the two system basically touch each other. In the case of nuclei and cold atom 
systems~\cite{Magierski:2017,Magierski:2022,Bulgac:2017} this is clearly not the case when two nuclei collide, and before the matter 
overlap between the reaction partners occurs, the pairing fields of the two partners already ``know'' about the presence of each other, 
neglecting for the sake of the argument the presence of the long-ranged Coulomb interaction between protons in the two nuclei.  

And this ``communication'' problem and exchange of ``information'' between spatially separated superfluid systems is even more 
complicated than the mere influence of the pairing field of one system on another system. 
For the sake of the argument imagine that two hypothetical nuclei with zero electric charge (in order to exclude long-range Coulomb interaction)
or two fermionic neutral superfluid cold atomic clouds are separated by a distance much larger 
than the average interparticle separation in each system or larger than the range of the interparticle interaction. Such systems are still able 
to ``communicate'' due to the fact that the  ${\text u}_k(\xi)$-quasi-particle wave function components are continuum wave functions. 
Thus any change in one system, due to its own quantum evolution due to the short-range interaction between the particles localized in that system, is 
``communicated'' via the  ${\text u}_k(\xi)$-quasi-particle wave function components to the other system, 
which in principle could be at the other end of the universe. 
One might argue that this is simply an artifact of mixing systems with different particle numbers in the HFB approximation. However, this argument 
cannot be valid for systems with a finite number of particles, where the probability to have a system with a very large number of particle is exponentially small and 
insufficient to bring in material contact the two systems, as can be easily shown by performing a particle number projection, see Section~\ref{sec:V}. 

Moreover, even after particle number projection of the HFB equations the anomalous densities and the corresponding pairing fields have tails much longer 
than the matter distribution of the two subsystems. It is not clear yet whether within a theoretical treatment of the pairing correlations, 
where the particle numbers are exactly conserved, the pairing field will cease to have longer tails than the regular mean field. Even after particle number projection 
the tails of the pairing fields are not affected, see Section~\ref{sec:V}.   Using the two type of generalized Bogoliubov quasi-particle described above one can address 
these questions at least within the HFB approximation after particle number projection. 

\section{Time-dependent equations} \label{sec:IV}

One can prove that using either the full set of 
the original quasi-particle wave functions ${\textrm u}_k, {\textrm v}_k$ or the canonical set $\tilde{\textrm u}_k, \tilde{\textrm v}_k$
by solving the corresponding  time-dependent evolution equations
\begin{equation}
\label{eq:eq_ufg}
i\hbar \frac{\partial}{\partial t} 
                    \left ( \begin{array}{c}  \tilde{\textrm u}_k \\ \tilde{\textrm v}_k \end{array} \right ) 
                  =\left ( \begin{array}{cc}  h  &\Delta\\
                                                         \Delta^\dagger & -h^*  \end{array} \right )                  
                        \left ( \begin{array}{c}  \tilde{\textrm u}_k \\ \tilde{\textrm v}_k \end{array} \right ), 
\end{equation}
one obtains the same normal $n(\xi,\zeta)$ and anomalous $\kappa(\xi,\zeta)$ density matrices. The explicit time-dependence 
was here suppressed. 

The main difference 
with the static case is that if one starts with a system of canonical quasi-particle states at any time $t$, at the next time step the new set of
quasi-particle states ceases to be canonical.  In practice this is not a problem if one uses at all times the full set of quasi-particle states. However, it is  known that if at any given time one chooses to represent the density matrices using canonical quasi-particle states, for a 
sufficient accuracy one can obtain their representation with a significantly reduced number of states. For example, in performing 
fission dynamics simulations of an actinide nucleus one needs to represent the quasi-particle states on a spatial lattice of 
typical size $N_xN_yN_z=30\times 30\times60$~\cite{Bulgac:2019c,Bulgac:2020,Shi:2020}, in which case the total number of proton and neutron quasi-particle states is 
$2\times 2\times N_xN_yN_z=216\,000$. At any time during the time evolution one can however introduce the set of canonical quasi-particle states 
and represent the same normal and anomalous densities with a comparable numerical accuracy with about 5\,000 states or less, see Ref.~\cite{Bulgac:2023} 
and also Section~\ref{sec:VI}.
As a matter of fact, in static calculations accurate solutions were obtained with a significantly reduced set of canonical wave 
functions~\cite{Chen:2022}, for example in the case of $^{240}$Pu these authors reproduced the static states binding energies with at
most 400 neutron orbitals and 300 proton orbitals, but see also Section~\ref{sec:VI}.

The main reason one needs a number of orbitals greater than the particle number  for static calculations is the need to reproduce 
the anomalous density $\kappa({\bm r},\sigma,{\bm r}',\sigma')$, which  in the case of a local pairing potential diverges as $1/|{\bm r}-{\bm r}'|$ 
in the limit when $|{\bm r}- {\bm r}'| \rightarrow 0$, as was shown in Ref.~\cite{Bulgac:1980}. This type of divergence is similar to the divergences one encounters 
in quantum field theory and they require a renormalization and regularization of the anomalous density, which was performed for the first time 
in Refs.~\cite{Bulgac:2002,Bulgac:2002a} and implemented in a more accurate manner later~\cite{Bulgac:2019c,Bulgac:2020,Shi:2020}. 
The physical reason why such a divergence appears is that for most fermionic superfluids one has 
a range of scales, from the range of interaction $r_0$, the average interparticle separation $1/\sqrt[3]{n}$, and the $s$-wave scattering length $a$ 
satisfying approximately the inequality
\begin{align}
r_0 \ll \frac{1}{\sqrt[3]{n}} \ll |a|
\end{align}  
 for example in dilute neutron matter in the neutron star crust or in cold fermionic atoms in the unitary regime when $r_0\rightarrow 0$ and 
 $|a|\rightarrow \infty$~\cite{Bulgac:2007,Tan:2008a,Tan:2008b,Tan:2008c,Bulgac:2011,Bulgac:2011a,Zwerger:2011,
 Braaten:2011,Castin:2011,Bulgac:2013a,Bulgac:2016x,Bulgac:2019}.  As \textcite{Tan:2008a,Tan:2008b,Tan:2008c} has proven 
 this singularity is present at all energies and all temperatures, irrespective of whether the system is superfluid or not, and in nuclei 
 this behavior is related to the short-range correlation between nucleons, a phenomenon known since the
 1950's~\cite{Levinger:1951,Levinger:1979,Levinger:2002,Tropiano:2022,Tropiano:2021} and observed in the last few years in experiments at 
 JLAB~\cite{Hen:2014,Hen:2017,Cruz-Torres:2018,Cruz-Torres:2021} and 
 studied theoretically by many~\cite{Frankfurt:1981,Frankfurt:1988, Piasetzky:2006,Sargsian:2005,Carlson:2015,Schiavilla:2007,Magierski:2022}.
 Upon regularization one can limit the sum over the quasi-particle wave functions to an energy interval close to the Fermi level, 
 and at the same time renormalize the strength of the pairing interaction, so as obtain the same pairing field $\Delta(\xi)$, 
 irrespective of the quasi-particle energy cutoff.
 
 The singularity of the anomalous density matrix $\kappa(\xi,\zeta)$ leads to a universal behavior 
 of the occupation probabilities of the canonical states at large momenta $n(p)\propto1/p^4$~\cite{Tan:2008a,Tan:2008b,Tan:2008c}, 
 which is also observed in simulations of nuclear systems, both in the static and time-dependent cases~\cite{Bulgac:2022,Bulgac:2022a,Bulgac:2022d,Bulgac:2023}.
 It is easy to check that the anomalous density matrix diverges in the same manner even in the BCS-approximation, when
 \begin{align}
 n_k = \frac{1}{2}\left  (1-\frac{ \varepsilon_k-\mu }{ \sqrt{ (\varepsilon_k-\mu)^2+\Delta^2} } \right )
 \end{align}
 as 
 \begin{align} 
 \Delta = -g \sum_k \sqrt{n_k(1-n_k)}, \label{eq:D}
 \end{align}
 where $\varepsilon_k$, $\mu$, $\Delta$ and $g$ are the  (canonical) single-particle energies, the chemical potential, the pairing gap, and the strength of the pairing interaction,
 unless the sum in  Eq.~\eqref{eq:D} is ``artificially'' cutoff and the strength of the pairing interaction $g$ is renormalized accordingly, 
 a procedure widely used in nuclear physics for decades~\cite{Ring:2004}, after the necessary ``excuses'' have been expressed, 
 such as the ``total energy of the ground state has converged.''

If one uses an incomplete set of quasi-particle wave functions the antisymmetry of the anomalous density $\kappa(\xi,\zeta)$ is lost and  one should enforce it by hand as follows, see also Eq.~\eqref{eq:Cooper},
\begin{align}
 \kappa({\bm r},\sigma, {\bm r},-\sigma) &\equiv-\kappa({\bm r},-\sigma, {\bm r},\sigma)=\langle\Phi|\psi({\bm r},\sigma)\psi({\bm r},-\sigma)|\Phi\rangle \nonumber \\
 &= \frac{1}{2} \sum_{\tau=\pm} \sum_{l=1}^{\Omega} \tilde{\text v}_{l,\tau}^*({\bm r},-\sigma) \tilde{\text u}_{l,\tau}({\bm r},\sigma) \nonumber \\
 &-\frac{1}{2} \sum_{\tau=\pm} \sum_{l=1}^{\Omega} \tilde{\text v}_{l,\tau}^*({\bm r},\sigma) \tilde{\text u}_{l,\tau}({\bm r},-\sigma), \label{eq:anm}
 \end{align}
where ${\bm r} = {\bm r}'$, since one needs only the spatial diagonal part of the anomalous density in typical simulations, 
or equivalently considering only local pairing fields.
Implementing this simple correction ensures that in simulations with an incomplete set of quasi-particle/canonical wave functions 
the total particle number is conserved, even though other quantities are not reproduced correctly, see Section~\ref{sec:VI}.

\section{Particle-number projection} \label{sec:V}

For particle number projections we need to evaluate density matrices~\cite{Bulgac:2021c}
\begin{align}
\tilde{n}(\xi,\xi'|\eta_0) &= \langle \Phi|\psi^\dagger(\xi')\psi(\xi)|\Phi(\eta_0)\rangle, \\
\tilde{ \kappa}(\xi,\xi'|\eta_0) &= \langle \Phi|\psi(\xi')\psi(\xi)|\Phi(\eta_0)\rangle, \\
\tilde{\overline{ \kappa}}(\xi,\xi'|\eta_0) &= \langle \Phi|\psi^\dagger (\xi)\psi^\dagger(\xi')|\Phi(\eta_0)\rangle,
\end{align}
where $|\Phi(\eta_0)\rangle = \exp(i\hat{N}\eta_0)|\Phi\rangle$, which in terms of canonical quasi-particle wave functions (QPWFs) 
\begin{align} 
\!\!\!\!\!\!\tilde{n}(\xi,\xi'|\eta_0)&=\langle \Phi|\Phi(\eta_0)\rangle  
        \sum_{k=1}^{2\Omega} \frac{\tilde{{\textrm v}}_k^*(\xi)\tilde{{\textrm v}}_k(\xi')e^{2i\eta_0}}{1+(e^{2i\eta_0}-1)n_k},\label{eq:NNF}\\
\!\!\!\!\!\!\tilde{\kappa}(\xi,\xi'|\eta_0)&=\langle \Phi|\Phi(\eta_0)\rangle  
        \sum_{k=1}^{2\Omega} \frac{\tilde{{\textrm v}}_k^*(\xi)\tilde{{\textrm u}}_k(\xi')e^{2i\eta_0}}{1+(e^{2i\eta_0}-1)n_k}, \label{eq:NNF1}\\
\!\!\!\!\!\!\tilde{\overline{ \kappa}}(\xi,\xi'|\eta_0)&=\langle \Phi|\Phi(\eta_0)\rangle  
        \sum_{k=1}^{2\Omega} \frac{\tilde{{\textrm v}}_k(\xi)\tilde{{\textrm u}}^*_k(\xi')}{1+(e^{2i\eta_0}-1)n_k}, \label{eq:NNF2}     
\end{align}
where $\Omega =N_xN_yN_z$ and with the overlap given by
\begin{align}
\!\!\!&\langle \Phi|\Phi(\eta_0)\rangle = \prod_{k=1}^\Omega [u_k^2+e^{2i\eta_0}v_k^2]\nonumber\\
&= \prod_{k=1}^\Omega [1+(e^{2i\eta_0}-1)n_k],\\
\!\!\!&|\Phi(\eta_0)\rangle = \prod_{\mu=1}^\Omega \left ( u_\mu+e^{2i\eta_0} v_\mu a^\dagger_\mu a^\dagger_{\overline{\mu}}\right ) |0\rangle = \nonumber \\
 \!\!\!&=\prod_{\mu=1}^\Omega \left [ u_\mu \exp \left ( 
 e^{2i\eta_0} \frac{v_\mu}{u_\mu}  a^\dagger_\mu a^\dagger_{\overline{\mu}} \label{equ:overlap}
 \right ) \right ]|0\rangle,
\end{align}
where $\eta_0$ is  a gauge angle.
The particle number distribution is given by $P(N)=\langle \Phi|\hat{P}|\Phi\rangle$ where the particle projector is 
\begin{align}
&\hat{P} = \int_{-\pi}^\pi \frac{d\eta_0}{2\pi} e^{i\eta_0 (\hat{N}-N)},\\
&P(N)=  \int_{-\pi}^{\pi}\frac{d\eta_0}{2\pi} e^{-iN\eta_0} \prod_{k=1}^\Omega [u_k^2+e^{2i\eta_0}v_k^2], \label{equ:N}\\
& \sum_NP(N) =1.
\end{align}

One can show that the $N$-particle number-projected many-body wave function is given by
\begin{align}
|\Phi^N\rangle = \frac{\prod_{\mu=1}^\Omega u_\mu}{\sqrt{P(N)}} \frac{1}{\sqrt{(N/2)!}}
\left ( \sum_{\mu =1}^\Omega \frac{v_\mu}{u_\mu} a_\mu^\dagger a^\dagger_{\overline{\mu}}\right )^{N/2}|0\rangle ,\label{equ:PhiN}
\end{align}
where 
\begin{align}
|\Psi\rangle = \frac{1}{\sqrt{ \sum_{\mu = 1}^\Omega \frac{v_\mu^2}{u_\mu^2} }}
\left ( \sum_{\mu =1}^\Omega \frac{v_\mu}{u_\mu} a_\mu^\dagger a^\dagger_{\overline{\mu}}\right )|0\rangle
\end{align}
is the wave function of a ``Cooper pair,'' also known in the literature as a geminal. The number projected wave function $|\Phi^N\rangle$ is a sum of 
linearly independent $N$-particle Slater  determinants. $|\Phi^N\rangle$ is a state of $N/2$ ``Cooper pairs'' $|\Psi\rangle$ with proper normalization. 

For an arbitrary operator $\hat{O}$
the particle number projected value in a state with exactly $N$ particles  is defined as~\cite{Bulgac:2021c}
\begin{align} 
&\langle\hat{O}^N\rangle = \langle \Phi^N|\hat{O}|\Phi^N\rangle
,\\
&  \langle \Phi |\hat{O}|\Phi\rangle = \sum_N P(N) \langle\hat{O}^N\rangle.
\end{align}
where the fact that the operator $\hat{P} = \int_{-\pi}^\pi \tfrac{d\eta_0}{2\pi} e^{i\eta_0 (\hat{N}-N)}$ is a projector was used.
The particle number projected number density matrix is
\begin{align}
& \tilde{n}(\xi,\xi' | N) = \frac{1}{P(N)}  \int_{-\pi}^\pi\frac{d \eta_0}{2\pi} e^{-i\eta_0N}   \tilde{n}(\xi,\xi'|\eta_0), \label{equ:nN} \\
& N = \int d\xi \, \tilde{n}(\xi,\xi|N),\\
& \sum_{k=1}^{2\Omega} {\text n}_k  = \sum_{N=1}^{2\Omega} N P(N),\\
& \sum_NP(N) \tilde{n}(\xi,\xi' | N)e^{i\eta N} = \tilde{n}(\xi,\xi' |\eta) 
\end{align}
and similar corresponding equations for the anomalous number densities $\tilde{\kappa}(\xi,\xi' | N), \tilde{\overline{\kappa}}(\xi,\xi' | N)$.

In order to evaluate various particle number projected number densities it is helpful to introduce an additional new 
set of particle number projected occupation probabilities
\begin{align}
\!\!\!\! \tilde{\text{n}}_k^N =  \int_{-\pi}^\pi\frac{d \eta_0}{2\pi} \frac{e^{-i\eta_0N}}{P(N)}
\frac{ \langle \Phi|\Phi(\eta_0)\rangle e^{2i\eta_0}}{u_k^2+e^{2i\eta_0}v_k^2}  \equiv \frac{P_k(N-2)}{P(N)},\label{equ:Nnk}\\
\!\!\!\! \tilde{{\text m}}_k^N =  \int_{-\pi}^\pi\frac{d \eta_0}{2\pi} \frac{e^{-i\eta_0N}}{P(N)} 
\frac{\langle \Phi|\Phi(\eta_0)\rangle }{u_k^2+e^{2i\eta_0}v_k^2} \equiv \frac{P_k(N)}{P(N)}, \label{equ:Ank}
\end{align}
with $\tilde{\text n}_k^N \geq 0$ and $\tilde{\text m}_k^N\geq 0$.
The particle number projected density matrices have then the form
\begin{align} 
\tilde{n}(\xi,\xi'|N)&=
        \sum_{k=1}^{2\Omega} \tilde{\text n}_k^N \tilde{{\textrm v}}_k^*(\xi)\tilde{{\textrm v}}_k(\xi'),\label{eq:mnf}\\
\tilde{\kappa}(\xi,\xi'|N)&=
        \sum_{k=1}^{2\Omega} \tilde{\text n}_k^N\tilde{{\textrm v}}_k^*(\xi)\tilde{{\textrm u}}_k(\xi'), \label{eq:mnf1}\\
\tilde{\overline{ \kappa}}(\xi,\xi'|N)&= 
        \sum_{k=1}^{2\Omega} \tilde{{\text m}}_k^N\tilde{{\textrm v}}_k(\xi)\tilde{{\textrm u}}^*_k(\xi'). \label{eq:mnf2}     
\end{align}
One can show that
\begin{align}
(1-{\text n}_k)\tilde{\text m}_k^N +{\text n}_k\tilde{\text n}_k^N =1,\quad 
N = \sum_k {\text n}_k\tilde{\text n}_k^N. 
\end{align}
For $\tilde{\text n}_k =\tilde{\text m}_k\equiv 1$  Eqs.~(\ref{eq:mnf}-\ref{eq:mnf2}) lead to the corresponding particle number unprojected number density matrices.
In Eqs.~(\ref{eq:mnf}-\ref{eq:mnf2})  the probabilities $P_k(N- 2), P_k(N)$ are evaluated as in Eq.~\eqref{equ:N},
but with a missing contribution from quasi-particle state $k$
\begin{align}
P_k(N)=  \int_{-\pi}^{\pi}\frac{d\eta_0}{\pi} e^{-iN\eta_0} \prod_{l\neq k}^\Omega [u_l^2+e^{2i\eta_0}v_l^2].\label{equ:N_k}
\end{align}
From these expressions one can write in straightforward manner the corresponding expressions for the particle number projected
normal number, kinetic, current, spin, spin-current and anomalous densities.
What is notable about Eqs.~(\ref{eq:mnf}-\ref{eq:mnf2}) is that the contribution of the troublesome self-interacting term is excluded in all 
these one-body densities and their use is free of the self-energy problem widely discussed in 
literature~\cite{Stringari:1978,Perdew:1981,Stoitsov:2007,Bender:2003,Dobaczewski:2007,Duguet:2009,Bender:2009,Lacroix:2009,Hupin:2011,Duguet:2015}. \textcite{Hupin:2011} 
have obtained similar, though somewhat more complicated,  relations.

A serious issue, which is not completely resolved within DFT and other mean field frameworks is the so called self-interaction 
often raised as a limitation of DFT for either normal and superfluid systems,
see Ref.~\cite{Sheikh:2021} for a recent review and many earlier references. 
Within Kohn-Sham framework~\cite{Kohn:1965} the exchange and correlation contribution to the energy density 
functional are ``parametrized'' on an equal footing as a functional or function of the number density, 
and for that reason it appears that the single particle contribution 
to the number density appears as a self-interaction for a system with one particle only in particular. 
In the case when only two-body interactions are present, in order to evaluate the total energy of the system one needs
to evaluate the two-body density matrix
\begin{align}
&{\text n}_2(\xi,\zeta,\zeta',\xi') = 
\langle \Phi|\psi^\dagger(\xi)\psi^\dagger(\zeta) \psi(\zeta')\psi(\xi')|\Phi\rangle  \nonumber \\
&= \frac{1}{2}[ {\text n}_1(\xi,\xi')    {\text n}_1(\zeta,\zeta')
                   - {\text n}_1(\xi,\zeta'){\text n}_1(\zeta,\xi') ]  \nonumber \\     
& +  {\text n}_{corr}(\xi,\zeta,\zeta',\xi'), \\  
&{\text n}_1(\xi,\zeta) = \langle \Phi|\psi^\dagger(\zeta)\psi(\xi)|\Phi\rangle,\\
& {\text n}_{corr}(\xi,\zeta,\zeta',\xi')=  \kappa(\xi,\zeta)\kappa^*(\xi',\zeta'), \label{eq:ncorr}
\end{align}
where ${\text n}_{corr} (\xi,\zeta,\zeta',\xi')$ here is given only in the case of a HFB many-body wave function $|\Phi\rangle$.
Clearly the two-body density matrix vanishes when either $\xi=\zeta$ or $\xi'=\zeta'$ by definition, and so does its explicit 
form in case of an HFB many-body wave function, see Eq.~\eqref{eq:anm}, which is 
used in the extension of DFT, in the spirit of the Kohn-Sham DFT approach, to superfluid systems, called the 
Superfluid Local Density Approximation (SLDA)~\cite{Bulgac:2007,Bulgac:2011a,Bulgac:2013,Bulgac:2019}. 
It is clear that self-interacting contribution arising from the pairing correlations to the correlation energy is absent. 

Therefore, the only source for a self-interacting energy contribution can appear from the parameterization of the exchange 
contribution in DFT, a problem which is absent when using 
the particle number projected number density. 
In the treatment of the unitary Fermi gas in Refs.~\cite{Bulgac:2007,Bulgac:2011a} a relatively large discrepancy 
was observed when comparing the Quantum Monte Carlo (QMC) results to the SLDA results only in the case of two 
fermions with opposite spins, where exchange is absent, but the self-interacting energy is present in the energy 
density functional, as the unprojected particle number density was used. 
For larger particle numbers the agreement between QMC and SLDA results was always within the QMC 
statistical errors.

With the particle number projected densities one can evaluate the energy of 
a nucleus for a fixed particle number within DFT, 
as in the ideology of DFT the energy is well defined once the energy density functional is known and the relevant 
number densities as well. In the mean field framework, theorists refer to this recipe as the projection either before 
or after energy variation. The difference between ``variation before'' or ``variation after'' recipes is only in the manner 
the quasi-particle wave functions are determined. In the ``after variation'' approach, one solves the unprojected mean field 
equations and determines the energy minimum and the particle and/or the angular and/or parity projection are 
performed after the minimum is found and the projected energy is calculated. In the projection-before-variation approach,  
one performs the variation on the projected energy functional at first and determines the minimum energy 
subsequently. In determining the next step in a self-consistent procedure one can use however the number 
projected densities instead of the particle number unprojected energies, and in principle arrive at the same result. 
We want to stress that particle number projection is very inexpensive to perform and as a result one can perform 
DFT calculation for superfluid systems with exact particle numbers and in this case the self-energy 
terms are entirely absent. In ``parameterizing''  the energy density function therefore one can treat independently 
the exchange and the pairing correlation energies, thus removing a difficulty with the theoretical treatment of such systems~\cite{Sheikh:2021}.

\section{Examples from fission simulations} \label{sec:VI}

\subsection{On the use of a reduced set of canonical wave functions} \label{ssec:VIA}

All our fission dynamics simulations were perform using a fully self-consistent set of quasi-particle wave functions on a spatial lattice 
$30^2\times60$ fm$^3$ with a lattice constant of $l=1$ fm, corresponding to a momentum cutoff in one direction 
$p_{cut}= \pi\hbar/l\approx 600$ MeV/c, corresponding to a nucleon kinetic energy of more $\approx$ 180 MeV 
(or more than 360 MeV kinetic energy in the center of mass for two colliding nucleons),
significantly larger than the Fermi energy of symmetric nuclear matter, and very close to the upper energy considered in chiral EFT approaches. 
The self-consistent equations were solved for the nuclear density functional SeaLL1~\cite{Bulgac:2018}, 
at first using the HFBTHO code ~\cite{Navarro:2017}, after which we ported the proton and neutron number, current, kinetic energy and anomalous 
densities to our 3D spatial lattice and continued the self-consistent procedure until full convergence. The size of the HFBTHO basis set 
for either proton or neutron system was about 8\,000 compared to the basis set of our 3D spatial lattice, which was 108\,000.  In Table~\ref{table:initial} we display 
some properties of the initial state at the top of the outer fission barrier, evaluated either with the full set of quasi-particle wave functions obtained 
in the self-consistent procedure described here, or using the entire set of canonical wave functions instead, or only a reduced set of 
canonical wave functions, selected with the largest occupation probabilities. As expected the last two entries in the Table~\ref{table:initial} 
are identical. It is important to compare however the initial converged energy obtained with the HFBTHO code  $-1,780.73$ MeV, 
which is considerably above the energy $-1,785.41$ MeV we obtain on the 3D lattice.

\begin{table}[h]
\centering
\caption{  In the first column we state the number of canonical wave functions with the largest occupation probabilities, 
for equal number of proton and neutron states, for a specific initial state near the outer fission barrier for $^{236}$U, 
used by us in fission dynamics simulations. In the last row we show the corresponding quantities evaluated with 
the full set of quasi-particle wave functions, obtained by solving the static self-consistent DFT equations with 
appropriate constraints~\cite{Shi:2020}. 
In  the subsequent columns we show the total energy (in MeV), total neutron and proton particle numbers, 
and total quadrupole (in b) and octupole (in b$^{3/2}$)  deformations evaluated with the corresponding set of quasi-particle wave functions. The calculations were 
performed on spatial lattice $\Omega= N_xN_yN_z=30^2\times60$ with a lattice constant  of 1 fm, and the size of the full set of quasi-particle wave function is 
$2\Omega = 2\times 30^2\times60 = 108\,000$, for either proton or neutron systems. For this simulation trajectory the initial state was given a 1.17 MeV 
energy boost for  the collective quadrupole modes~\cite{Shi:2020}.
All densities were evaluated with either the usual quasi-particle or the canonical wave functions, as specified in the first column, at the beginning of the simulation.}
\begin{tabular}{cccccc}
\hline \hline
No. States & Energy & N & Z & $Q_{20} $ & $Q_{30}$ \\ \hline
500 &  -1779.19 & 143.94 & 91.99 & 169.73 & 22.29   \\ 
1\,000& -1780.64& 143.97& 92.00& 169.76& 22.29  \\ 
2\,000& -1782.64&143.99& 92.00& 169.78& 22.29 \\ 
5\,000&  -1785.31& 144.01& 92.00& 169.79& 22.30 \\ 
50\,000&  -1785.41& 144.01& 92.00& 169.79& 22.30 \\ 
108\,000 (Canonical)&  -1785.41& 144.01& 92.00& 169.79& 22.30 \\ 
108\,000 (Standard)&-1785.41& 144.01& 92.00& 169.79& 22.30\\ \hline \hline
\end{tabular}

\label{table:initial}
\end{table}

\begin{table}[h]
\centering
\caption{   In the fission run performed with initial set being the entire set of canonical wave functions as in table~\ref{table:initial}, 
we have evaluated the final proton and neutron numbers in the final states, but using only the states $500, \ldots, 50\,000$ 
with  the largest initial canonical occupation probabilities, using initial conditions listed in Table~\ref{table:initial} with boosting.    }

\begin{tabular}{ccccccc}
\hline \hline
No. States & $Z_i$ & $Z_f$ & $\Delta Z$ & $N_i$ & $N_f$ & $\Delta N$ \\ \hline 
500 & 91.99 &   91.87 &    0.12  &  143.94  &   141.74    &    2.20   \\ 
1\,000 & 92.00  &     91.89   &  0.11  &   143.97    &   141.86   &    2.11 \\ 
 2\,000  &  92.00   &    91.90  &    0.11 &      143.99    &   141.95  &     2.04\\ 
5\,000   &    92.00     &  91.91   &  0.09 &      144.01   &    142.11   &    1.90 \\ 
50\,000    &   92.00 &      92.00   &  0.01    &   144.01   &    143.48   &    0.52 \\ 
108\,000 &  92.00   &    92.00 &  $1.0\times 10^{-6}$ &      144.01   &    144.01  & $2.9\times 10^{-5}$ \\ \hline \hline
\end{tabular}

\label{table:PN}
\end{table}

\begin{table}[h]
\centering
\caption{Initial and final state convergence, same as Table \ref{table:PN}, with initial $Q_{20}=140.02$ b and $Q_{30}= 14.63$ b$^{3/2}$.}
\begin{tabular}{ccccccc}
\hline\hline
No. States & $Z_i$ & $Z_f$ & $\Delta Z$ & $N_i$ & $N_f$ & $\Delta N$ \\ \hline
500 & 91.97 &   91.61 &     0.36   &  143.96  &   139.74    &    4.21   \\ 
1\,000 & 91.98  &     91.64   &  0.34  &   144.00    &    139.92    &     4.07 \\ 
 2\,000  &  91.99   &    91.66   &    0.33 &      144.02    &   140.06  &     3.96\\ 
5\,000   &    91.99     &  91.69   &  0.30 &      144.03   &    140.33   &    3.70 \\ 
50\,000    &   91.99 &      91.97   &  0.017    &   144.03   &    143.23   &    0.80 \\ 
108\,000 & 91.99   &    91.99 &  $6.5\times 10^{-6}$ &      144.03   &    144.03  & $1.1\times 10^{-4}$ \\ \hline \hline
\end{tabular}

\label{your-table-label}
\end{table}

On the other hand the total initial energy, particle numbers, and deformations are reproduced adequately only when at least 
5\,000 canonical wave functions are taken into account. One might have naively assumed that using 500 canonical wave functions for 
both proton and neutron systems the properties of the initial state would be quite well reproduced, as \textcite{Chen:2022} expected. 
These authors used used 400 and 300 canonical wave functions for the neutron and proton systems in order to evaluate 
the energy of the fission isomer  $^{240}$Pu, which should be compared with our claim that at least 5\,000 canonical 
wave functions are needed  to obtain a comparable accuracy for the total energy of a similar nucleus. Reducing the number of 
canonical wave functions from 5\,000 to 2\,000, see Table~\ref{table:initial}, leads to a significant error in the energy.
In Table~\ref{table:PN} we have evaluated the final proton and neutron numbers, where we used the entire set of canonical wave 
functions to evolve the system, see penultimate row in Table~\ref{table:initial}, within a reduced set of quasi-particle 
wave functions with the corresponding largest initial occupation probabilities, in the range from 500 to 50\,000, 
for both neutron and proton subsystems. This demonstrates that the naive choice of a  reduced set initial quasi-particle wave functions, 
which however reproduced  accurately the initial neutron and proton numbers cannot reproduce the corresponding 
final total neutron and proton numbers. 

It is important to appreciate that in the results reported in this subsection, we have always performed the simulations with either the full set of quasiparticle wave functions or equivalently with the full set of canonical wave functions, and we have always obtained perfect agreement between these runs, as expected. The point we are making is that if one assumes that only a reduced set of canonical wave functions would be numerically accurate, simply selecting a reduced set of canonical wave functions with initially largest occupation probabilities $n_k(0)$ in evaluating the properties of the final state one does not obtain correct results, see also Table \ref{your-table-label}. We will come back to this aspect from a different point of view in the subsection~\ref{ssec:VIC}.   

A feature we generally observe in our time-dependent simulations is that the quality of the initial state plays a significant role in the accuracy of the 
entire simulation until the fission fragments (FFs) are fully separated. Better converged solutions for the initial state lead to significant increase in the 
accuracy of the simulations, judged by either the conservation of the particle numbers or total energy. Comparing the results obtained 
with either the full set of quasi-particle wave functions, last row in Table~\ref{table:initial}, with the full set of the canonical wave functions 
we observed relative errors at the level of $10^{-4\ldots -8}$, after performing about $30\,000$ time-steps, depending on the quantity.  
The total particle number is conserved at the level of $10^{-7\ldots -8}$ and the total energy at the level of noticeably less than 1 KeV, 
thus a relative  error less than $10^{-6}$.

\begin{figure}[h]
\includegraphics[width=1.0\columnwidth]{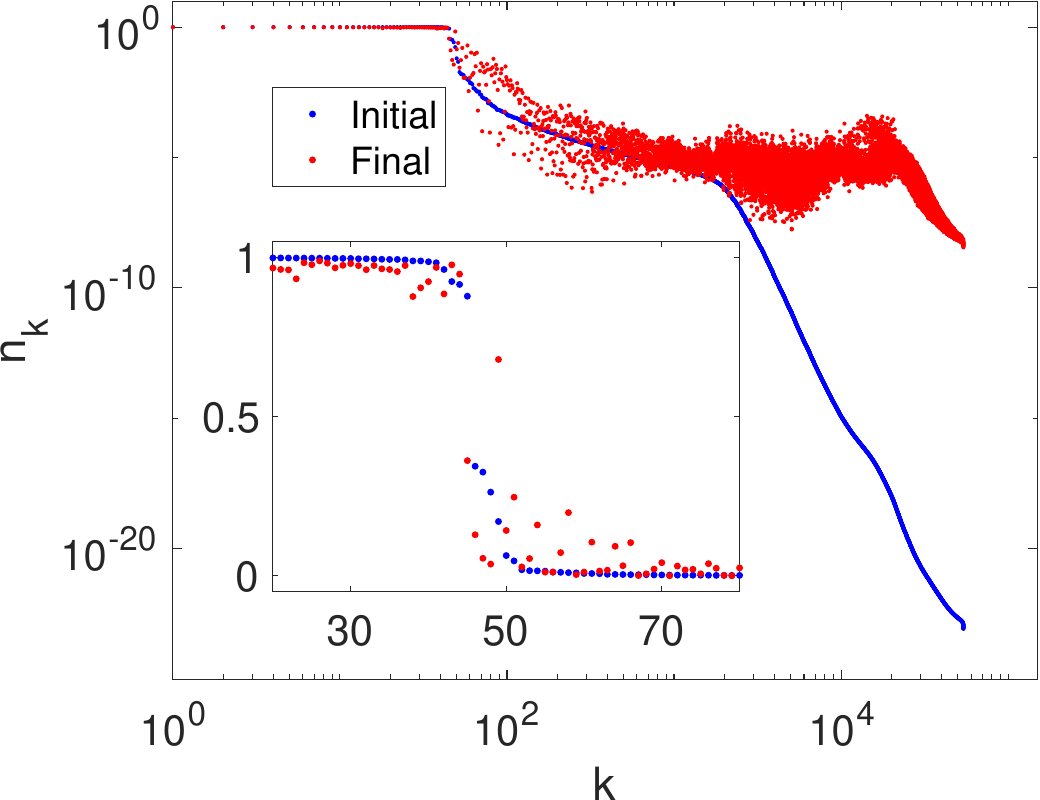}  
\includegraphics[width=1.0\columnwidth]{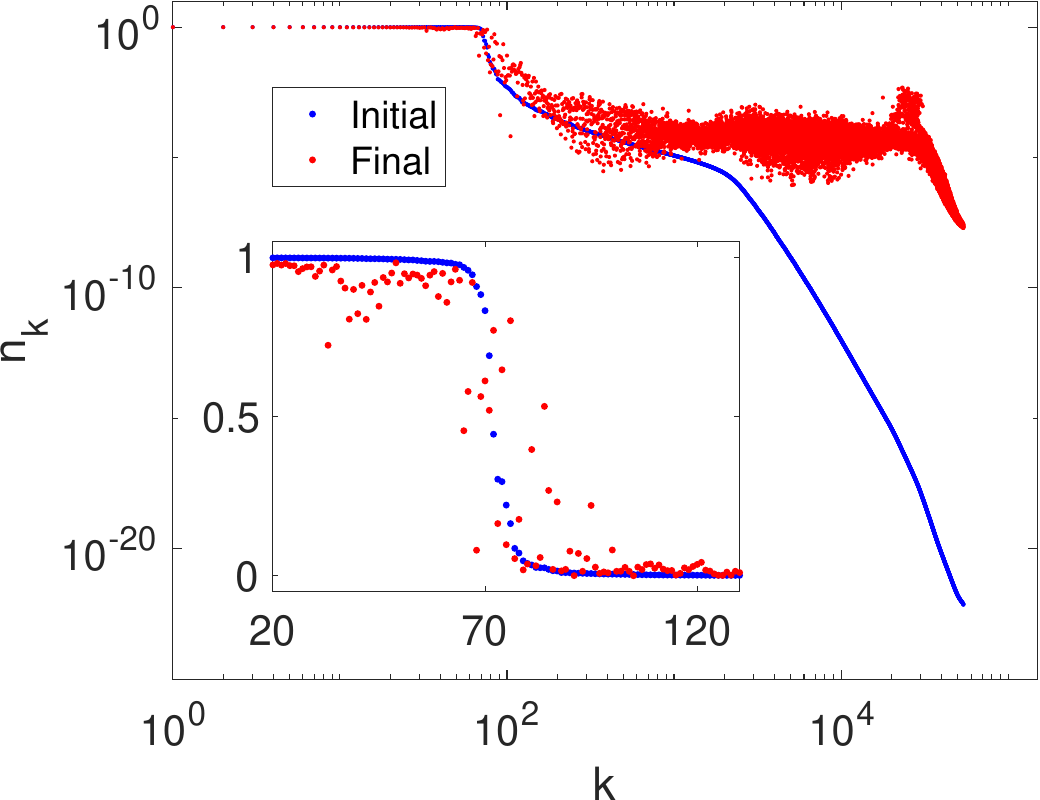}  
\caption{ \label{fig:PartNumb} 
 In the upper and lower panels we show the initial and final proton and neutron single-particle occupation probabilities, 
 evaluated performing a fission simulation of $^{236}$U with the NEDF SeaLL1, starting 
 near the top of the outer fission barrier, with the full set of initial quasi-particle wave functions, see Table~\ref{your-table-label},  
 until the FFs are  partially fully separated. Since the occupation probabilities are double degenerate, we display only 1/2 of their spectrum, in this case only $\Omega=54\,000$.
 The results presented in this figure 
 were obtained with $Q_{20}=140.02$ b and $Q_{30}= 14.63$b$^{3/2}$ set of initial conditions.}
\end{figure}  


\begin{figure}
\includegraphics[width=1.0\columnwidth]{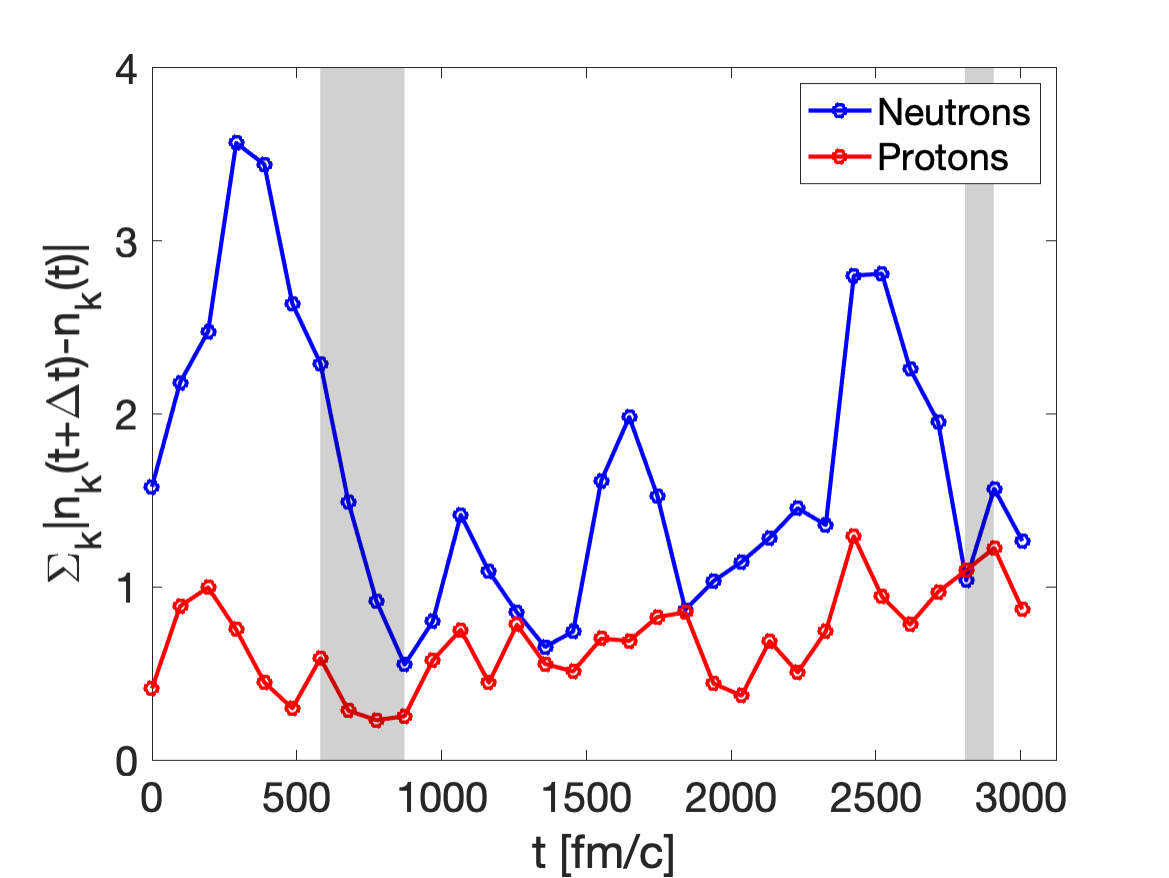}  
\caption{ \label{fig:nkdiff1}  
Total absolute difference between single-particle occupation probabilities at time $t$ and time $t+\Delta t$, where $\Delta t$ = 97 fm/c. The two shaded regions indicate when the neck is first formed between the fission fragments (left) and when it ruptures (right). Initial conditions are the same as in Fig.~\ref{fig:PartNumb}.
}
\end{figure}

\begin{figure}
\includegraphics[width=1.0\columnwidth]{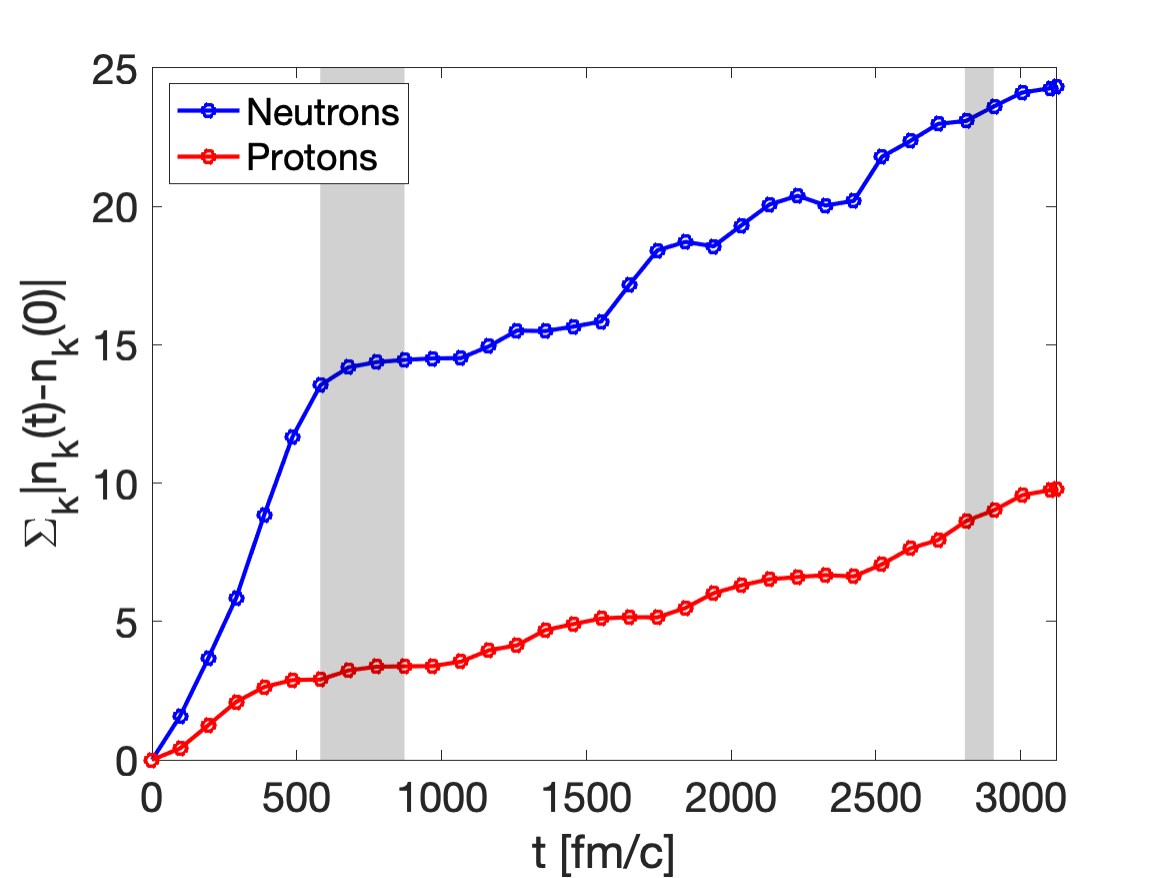}  
\caption{ \label{fig:nkdiff2}  
The sum of the absolute differences between the single-particle occupation probabilities at the initial time and at time $t$. Initial conditions are the same as in Fig.~\ref{fig:PartNumb}.
}
\end{figure}

\subsection{Non-Markovian character of fission dynamics}\label{ssec:VIB}

In Fig.~\ref{fig:PartNumb} we compare the initial canonical occupation probabilities $n_k(0)$ for both neutrons and protons, 
which change by more than 20 orders of magnitude, numbers we claim are numerically accurate, see also Ref.~\cite{Bulgac:2023}, 
with the final occupation probabilities  $n_k(t)=\sumint d\xi|{\text v}_k(\xi,t)|^2$, obtained from the time evolved ${\text v}_k$-components of the quasiparticle wave functions.
As we mentioned above and in Refs.~\cite{Bulgac:2023,Bulgac:2022,Bulgac:2022a,Bulgac:2022d}, even if one starts with a set of canonical wave functions, 
at the next time-step these wave functions fail to remain canonical. This is understandable, as in a framework where particle 
collisions are allowed, and they are allowed when pairing is taken into account beyond the static BCS approximation, 
the single-particle occupation probabilities  change~\cite{Bertsch:1980,Bertsch:1987,Bertsch:1991,Bertsch:1994,Bertsch:1997,Stetcu:2011,Bulgac:2019c,Bulgac:2020,Bulgac:2023} 
and the canonical occupation probabilities change in time, which means the entropy of the system changes. See the discussion of Figs.~\ref{fig:entropy}-\ref{fig:Occup} below.

\begin{figure}[h]
\includegraphics[width=0.3235\columnwidth]{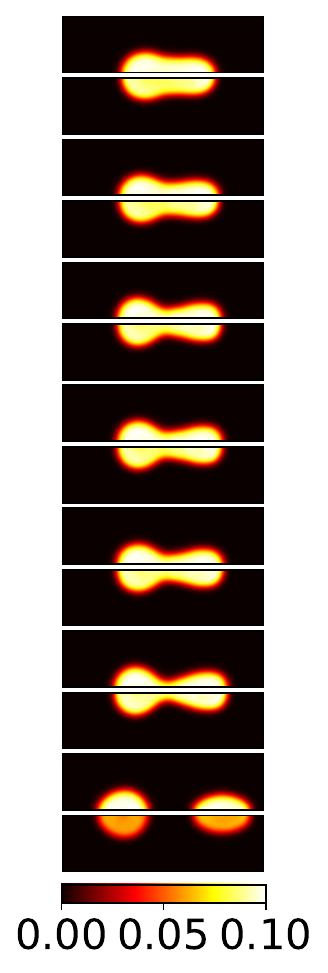}  
\includegraphics[width=0.3\columnwidth]{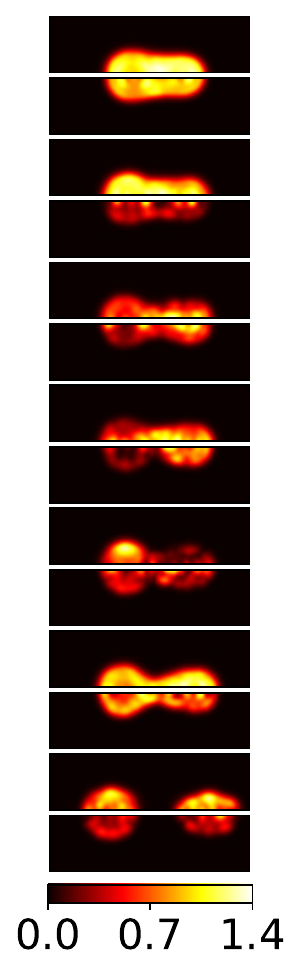} 
\caption{ \label{fig:DensEvol}  
Time frames illustrating the profiles of the neutron (top) and proton number densities (first column) and the absolute value so the pairing gap (second column), separated in time by about 500 fm/c.  The last row are for colorbars for the densities in units of  fm$^{-3}$ and MeV for the pairing gaps. The initial state was the same as in Fig.~\ref{fig:PartNumb}.}
\end{figure}  

In Figs.~\ref{fig:nkdiff1} and \ref{fig:nkdiff2} we plot the sum of the absolute changes in the single-particle occupation probabilities $n_k(t)$  at some fixed time intervals $\Delta t= 97$ fm/c and the absolute differences between the initial and time dependent  single-particle occupation probabilities. Similar results have been reported in Ref.~\cite{Bulgac:2022}, however, with slightly different initial conditions for the same nucleus $^{236}$U with the same NEDF SeaLL1, but for smaller values of $\Delta t = 37$ fm/c. In that case the initial quasiparticle wave functions were obtained using the code HFBTHO~\cite{Navarro:2017}, placed on the 3D spatial lattice, and only adjusting the proton and neutron chemical potentials to fix the correct average particle numbers. In all the simulations reported here we have run the static SLDA code on the 3D spatial lattice until full self-consistency was achieved. Since the phase space is much larger 
on a 3D spatial lattice than that used in HFBTHO code and since the kinetic energy and anomalous densities are formally diverging in a 3D space~\cite{Bulgac:1980}, the self-consistent SLDA equations need to be regularized and renormalized~\cite{Bulgac:2002,Shi:2020}. One important benefit of performing this additional self-consistency evaluation of the SLDA equations on the 3D spatial lattice is a much more numerically accurate solution of the TDSLDA equations.   

These results demonstrate that during the entire fission dynamics, even after full the FF spatial separation, these occupation probabilities evolve with time, as expected for a non-equilibrium process of an isolated quantum system. These results provide a direct confirmation of the mechanism envisioned by Bertsch~\cite{Bertsch:1980,Bertsch:1987,Barranco:1990,Barranco:1988,Bertsch:1997,Bulgac:2022}, describing how nuclei experience shape changes through the redistribution of the single-particle occupation probabilities, facilitated by the pairing correlations and the correct implementation of the hydrodynamic continuity equation.  
The mechanism for nuclear shape change advocated by Bertsch implies that the single-particle occupation probabilities change through
independent jumps at the single-particle level crossings and that these jumps are uncorrelated~\cite{Bertsch:1980,Bertsch:1987,Barranco:1990,Barranco:1988,Bertsch:1997}.

While the total particle number $\sum_k n_k(t)$ is conserved during the time evolution,  the individual single-particle occupation probabilities change significantly, and one might be naively led to assume that the change of a particular occupation probability $0\le n_k(t)\le  1$ is random. If that would be the case, exactly as in the case of Brownian motion
one would expect that
\begin{align}
\sigma_2(t) = \sqrt{\sum_k (n_k(t)-n_k(0))^2} \propto \sqrt{t}, \label{eq:sig2}
\end{align}
while the obviously, at large times, the closely related quantity $\sigma_1(t)$ illustrated in Fig.~\ref{fig:nkdiff2} 
\begin{align}
\sigma_1(t) = \sum_k |n_k(t)-n_k(0)| \propto t, \label{eq:sig1}
\end{align}
which can be characterized as a ``ballistic'' behavior of the single-particle occupation probabilities $n_k(t)$, patently a non-stochastic and therefore non-Markovian behavior. 
Since we are solving quantum equations for quasiparticle wave functions the case can be made that instead of the quantity $\sigma_1(t)$ one should instead consider 
\begin{align}
    \sigma_0(t) = \sum_k \left |\sqrt{n_k(t)} -\sqrt{n_k(0)}\right |,
\end{align}
since $\sqrt{n_k(t)}$ is proportional to the amplitude of each quasiparticle wave function ${\text v}_k(\xi,t)$, since these wave functions are the actual variables in the dynamic equations, similarly to the coordinates in the Langevin equation for example. The phases of the quasiparticle wave functions carry information about the currents, which to some extent were previously characterized by us when defining the collective flow kinetic energy~\cite{Bulgac:2019c,Bulgac:2020}
\begin{align}
E_{coll}(t)=\int d{\bm r} \frac{ m n({\bm r},t){\bm v}^2({\bm r},t)}{2},
\end{align}
where ${\bm v}({\bm r},t)$ is the hydrodynamic velocity of the nuclear matter. Unfortunately this quantity  also includes after scission the accumulated FF kinetic energy due to their Coulomb repulsion, known as total kinetic energy of the FFs. 

 The quantum mechanical nature of the nuclear shape proves however to be more complex than envisioned by Bertsch~\cite{Bertsch:1980,Bertsch:1987,Barranco:1990,Barranco:1988,Bertsch:1997,Bulgac:2022} and these jumps appear to be highly correlated in time, 
which is a qualitatively new aspect of fission dynamics in particular. This aspect is particularly interesting, since as we have proven in earlier fission simulations~\cite{Bulgac:2016,Bulgac:2019c,Bulgac:2020}, the descent from the top of the outer barrier to the scission configuration is a highly dissipative process, in which case one would expect that stochasticity of the dynamics might play a crucial role. In the presence of a strong dissipation, stochasticity in case of classical dynamics and in numerous phenomenological fission models is modelled with a Langevin force~\cite{Vogt:2009,Vogt:2013,Verbeke:2018,
Albertsson:2020,Ishizuka:2017,Randrup:2011,Randrup:2011a,
Randrup:2013,Sierk:2017,Moller:2001,Moller:2004,Litaize:2012,Becker:2013}. This is a qualitatively new situation in non-equilibrium dynamics, so far never discussed in literature as far as we can judge, where in the presence of strong dissipation, memory effects are also very strong and the single-particle occupation probabilities dynamics show a clear non-Markovian behavior.  

The results in Figs.~\ref{fig:nkdiff1}-\ref{fig:nkdiff2} show that the single-particle occupation probabilities, illustrated there at time intervals separated by $\Delta t = 97$ fm/c change rather in a continuous manner and not as individual jumps. A jump at a ``Landau-Zenner level crossing'' is not instantaneous, but is coherently coupled with other jumps, which occur before or after a particular level crossing and that leads to a rather strong quantum coherence.  In the end the change in the quantity $\sigma_1(t)$ instead of being random has a rather a well defined directed evolution, towards the equilibration  of the quantum many-body system.

The quantity $\sigma_1(t)$ appears to change at two different rates for times smaller than  500-600 fm/c and at a slower rather for larger times. 
We see three different sources for this behavior. i) With time the strength of the pairing correlations and the absolute magnitude of the pairing  gaps decreases, though it does never vanish, see Fig. \ref{fig:DensEvol} and ~\cite{Bulgac:2016,Bulgac:2019c,Bulgac:2020}. 
ii) The fissioning nucleus and the FFs after separation still convert deformation energy into thermal energy, which leads to increasing occupation probabilities of higher energy single-particle states. iii) 
The neck appears to start forming at times 500-700 fm/c~\cite{Bulgac:2019c,Bulgac:2022,Bulgac:2023} and scission occurs at time 2\,300-2\,700 fm/c, depending on the initial conditions considered.
In all TDDFT simulations one has to choose the initial nuclear shape but imposing a shape constraint and requiring that the total energy is near the outer fission barrier. The shape constraint is not inherent in the many-body Schr\"odinger equation and as in the initial \textcite{Bohr:1939} paper this is merely a theoretical tool used since 1939. The nucleus during its descent from the top of the fission barrier needs to adjust to the absence of the artificial shape constraint imposed on the initial state. 
For times larger than 500-600 fm/c the nucleus appears to have settled to a different slower, but rather well defined rate,  a bit higher for the neutron system than for the proton system. The neck, depending on its size, increasingly impedes matter, linear and angular momentum, and energy exchange between the two emerging FFs, and that is another reason why the rate of ``equilibration'' we see in Fig.~\ref{fig:nkdiff2} settles to a smaller value. At the same time, scission, which occurs at much later times, between 2\,300-2\,700 fm/c depending on the initial conditions and NEDF used, does not appear to affect this rate.


Ever since \textcite{Boltzmann:1872} introduced the classical kinetic equations, and later on with their extension to quantum phenomena by \textcite{Nordheim:1928} and \textcite{Uehling:1933}, it was assumed that two-body collisions lead to a Markovian behavior of the many-body system, and thus to an absence of memory effects, similar to the case of Brownian motion of a single particle. The classical \textcite{Boltzmann:1872} and the quantum extension of the collision integral due to \textcite{Nordheim:1928} and \textcite{Uehling:1933} were stochastic in character. On the other hand, the extension of the TDDFT framework to superfluid phenomena is an extension of the time-dependent mean field  dynamics to include a particular kind of collision term relevant in superfluids~\cite{Bulgac:2022}, as the action of the pairing field $\Delta(\xi,t)$ on the quasiparticle wave functions ${\text u}_k(\xi,t), {\text v}_k(\xi,t)$, see Eq.~\eqref{eq:eq_ufg}, 
which is not stochastic. The effect of this ``quantum collision integral''  is equivalent to the action of the collision term on the phase space occupation probabilities $f_k(q,p,t)$ in the Boltzmann-Nordheim (BN)~\cite{Nordheim:1928} or Boltzmann-Uehling-Uhlenbeck (BUU)~\cite{Uehling:1933} kinetic equations. There is however  a major difference: while in the ``quantum'' BN/BUU equations one operates with occupation probabilities, the 
TDSLDA equations~\cite{Bulgac:2011,Bulgac:2011a,Bulgac:2013a,Bulgac:2019,Bulgac:2022} are formulated in terms of the quasiparticle wave functions, as expected in the case of a genuine quantum many-body framework. The consequences are quite fundamental, as TDSLDA can describe dynamical evolution of fermionic superfluids, in particular non-equilibrium dynamics; quantum turbulence; generation, life, dynamics,  and decay of quantum vortices; entanglement; and nontrivial aspects of nuclear collisions~\cite{Bulgac:2011,Bulgac:2016x,Magierski:2017,Bulgac:2017,Pecak:2021,Magierski:2022,Hossain:2022}, which are not accessible within a BN/BUU framework, since quantum interference and superpositions are not incorporated in the Boltzmann collision integral, either in its original classical form or that of Refs.~\cite{Nordheim:1928,Uehling:1933}.
        
\subsection{Particle-number projection and use of a reduced set of canonical wave functions in time-dependent simulations}\label{ssec:VIC}


\begin{figure}[h]
\includegraphics[width=1.0\columnwidth]{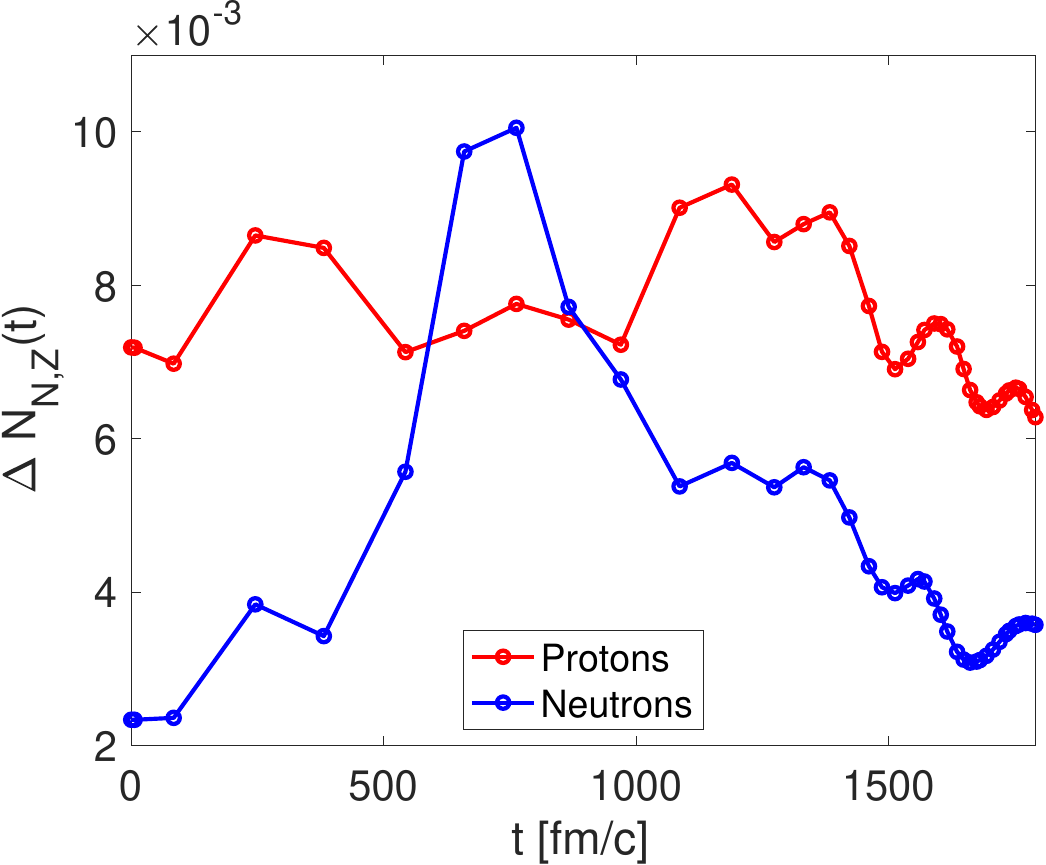}  
\caption{ \label{fig:ProjDensProj}  
The standard deviation of the integrated proton and neutron particle number projected densities $\Delta N_{Z,N}(t)$, defined in Eq.~\eqref{eq:dN}, as a function of time. Somewhat surprisingly the differences between 
the integrated particle number projected and unprojected number densities are very small, remembering that both $N$ and $Z$ are of ${\cal O}(100)$, 
thus the relative variance is at the level of ${\cal O}(10^{-4})$. The initial state was the same as in Table~\ref{table:initial}, but without the quadrupole collective energy boost.}
\end{figure}  

We have estimated average discrepancies between the particle-number projected and unprojected 
number densities obtained within TDSLDA at various times
\begin{align}
\Delta N_{N,Z}(t) =  \sqrt{\sumint_\xi  (n(\xi,\xi,t)-\tilde{n}(\xi,\xi,t|N))^2} \label{eq:dN}
\end{align}
and obtained  a very similar relative trend, see Fig \ref{fig:ProjDensProj}. 

In Ref.~\cite{Bulgac:2023} we have established that the character of the canonical wave functions depends on the spatial resolution adopted 
in the numerical treatment, thus on the lattice constant $l$. The canonical wave functions with non-negligible occupation probabilities are localized in the region where 
the matter distribution of the system is non-vanishing and their number is typically much larger than the number of particles in the system, 
but significantly smaller than the size of the entire set, which is $2\Omega = 2\times N_xN_yN_z$ for one type of nucleons. The spatial support 
for the canonical wave functions with negligible occupation probabilities is localized outside 
the region where matter distribution is localized.  Obviously, the border between the two regions is not sharp. In this work we report more accurate 
estimates for the minimal number of canonical wave functions needed in order to obtain accurate enough numerical solutions  
in the case of heavy nuclei than in Ref.~\cite{Bulgac:2023}.

\begin{figure}[h]
\includegraphics[width=1.0\columnwidth]{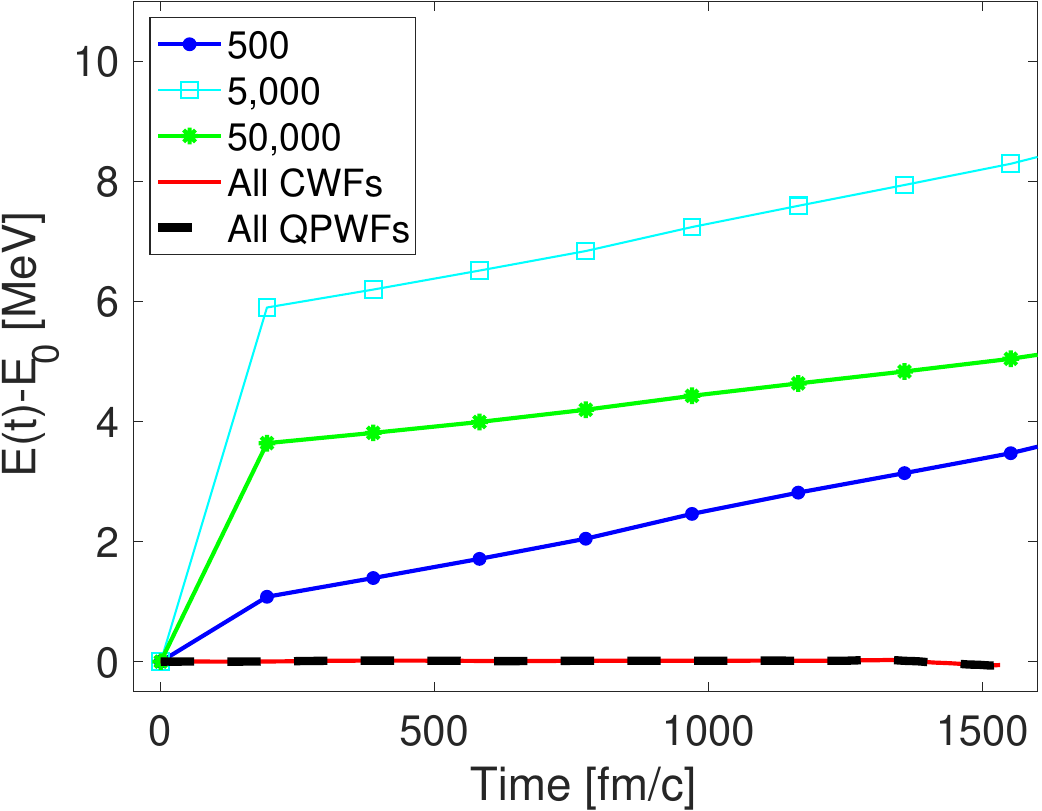}  
\includegraphics[width=1.0\columnwidth]{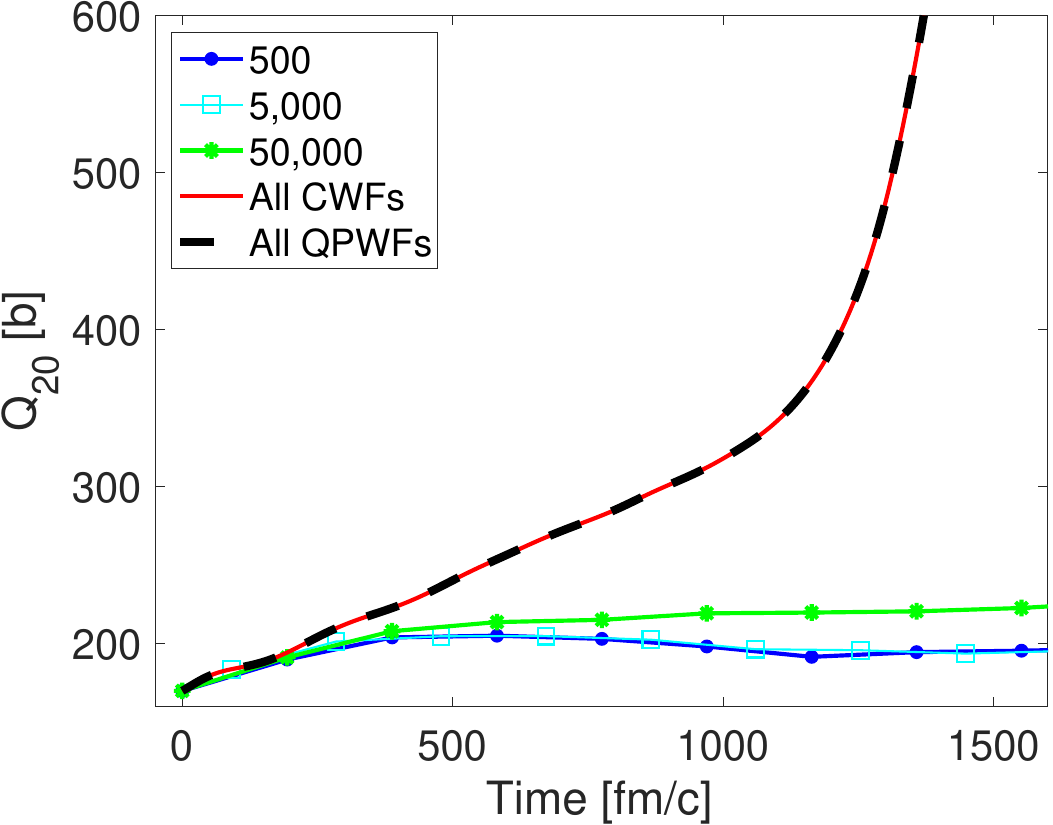}
\caption{ \label{fig:tot_en}  
Plots of the time-dependence of the total energy of $^{236}$U,  which theoretically should be conserved, as well as the time-dependence of the quadrupole moment, when various cutoffs are imposed on the initial set of canonical wave functions. 
The simulation with either the full set of quasi-particle or canonical wave functions are visually indistinguishable
and in this case the total energy is conserved.
These are results of simulation obtained with the set of initial deformation listed in Table~\ref{table:initial}.
}
\end{figure}

In Fig.~\ref{fig:tot_en} we show the total energy $^{236}$U as a function of time, depending 
on various numbers of canonical wave functions used as an initial set. With the exception of the simulations with 
the entire set of quasi-particle wave functions and the entire set of canonical wave functions used as initial 
conditions, which are indistinguishable on this plot, every run performed with
any number of wave functions as large as 50\,000 initial canonical wave functions, failed to lead to fission, 
and moreover the total energy of the system is not conserved, the $Q_{20}$ of the entire system basically does not change in time,
and the nucleus only keeps ``heating up.'' 
For any set of initial conditions, using the canonical wave functions with largest occupation  probabilities in the
 interval $500,\ldots,50\,000$, the nucleus does not fission, but only heats up. In the lower panel we display 
 the behavior of the quadruple moment of the entire nucleus $Q_{20}(t)$ as a function of time for the same 
 choice of initial conditions as for the upper panel. Only for the case when the entire set of  quasi-particle or 
 canonical wave functions is used the nucleus fissions, otherwise its size remains basically unchanged as a function of time.
 In this case the initial condition had an additional excitation energy of about 1.17 MeV and the simulations performed 
 with a reduced set of canonical wave function still did not fission.

 In Fig.~\ref{fig:tot_en},  
we report simulations performed with various limited sets of initial canonical quasi-particle wave functions. In past simulations, when we used a spherical cutoff for the 
pairing~\cite{Bulgac:2002,Bulgac:2002a,Bulgac:2016}, which is extremely well suited for static calculations, we often observed that during time-evolution the  system eventually failed numerically (overflow). Only after we gave up on using a spherical cutoff  
and implemented  the entire spectrum~\cite{Shi:2020} were we able to consistently obtain well behaved numerical solutions. It is 
remarkable to see that canonical states with an occupation number less that $10^{-10}$ acquire a significant 
occupation probability during the time-evolution.  Furthermore, we have shown that the naive expectation that a reduced set of canonical quasi-particle 
wave functions, which reproduce nuclear properties of the initial state with very good accuracy, can be used to study the dynamics of fission, is incorrect. 

The vast majority of simulations performed by other authors use the  TDHF+BCS or TDHF+TDBCS approximation~\cite{Zhang:2023,Li:2023a,Li:2023b,Ren:2022,Ren:2022a,Bender:2020,Scamps:2015,Scamps:2018,Scamps:2012}, where TD stands for time-dependent and HF for Hartree-Fock.
Either the BCS or TDBCS approximations are further approximations to the TD Hartree-Fock-Bogoliubov (TDHFB) equations with cutoffs in the number 
of levels allowed to participate in pairing.  Moreover, both BCS and TDBCS assume the spatial profiles of the $|{\text u}_k|^2$ and 
$|{\text v}_k|^2 $-components of the quasi-particle wave functions are identical, 
as in the initial BCS approximation~\cite{Bardeen:1957} for weak pairing correlations.  Even further, 
 the continuity equation is violated in the TDHF+TDBCS approximation~\cite{Scamps:2012}, which is still widely used today despite this~\cite{Zhang:2023,Li:2023a,Li:2023b,Ren:2022,Ren:2022a,Bender:2020,Scamps:2015,Scamps:2018,Scamps:2012} (to list a few studies).  In Ref.~\cite{Scamps:2019} a TDHFB implementation in a reduced basis set was used, which according to our present analysis might be problematic.
 
Both BCS and TDBCS represent more limited approximations than evolving a truncated set of canonical quasi-particle wave functions, or using a spherical cutoff, as was originally done in~\cite{Bulgac:2016}.  As described above, all trajectories performed in the canonical basis with a cutoff didn't fission. This is likely the reason why BCS and TDBCS simulations of (induced) fission~\cite{Zhang:2023,Li:2023a,Li:2023b,Ren:2022,Ren:2022a,
Bender:2020,Scamps:2015,Scamps:2018,Scamps:2012}
start with initial compound nuclei on the potential energy surface of the fissioning nucleus 
that are well below the saddle point state of the nucleus considered. Only in such situations,  for configurations where the Coulomb repulsion considerably exceeds the nuclear surface energy, could such simulations produce separated fission fragments. 

As \textcite{Meitner:1939}
correctly suggested, the nucleus during fission behaves like a liquid drop. Fission is due to the competition between the Coulomb 
and surface energies, and a correct hydrodynamic description of the nuclear shape dynamics is crucial, both at the classical and quantum level. 
How to achieve the correct description in the case of nuclei was not clear until Bertsch~\cite{Bertsch:1980,Bertsch:1987,Bertsch:1991,Bertsch:1997}  
identified the crucial role played by the pairing dynamics in the shape evolution of a fissioning nucleus, which was confirmed 
in 2016 in the first correct implementation of pairing dynamics for this process~\cite{Bulgac:2016}. 

\begin{figure}
\includegraphics[width=1.0\columnwidth]{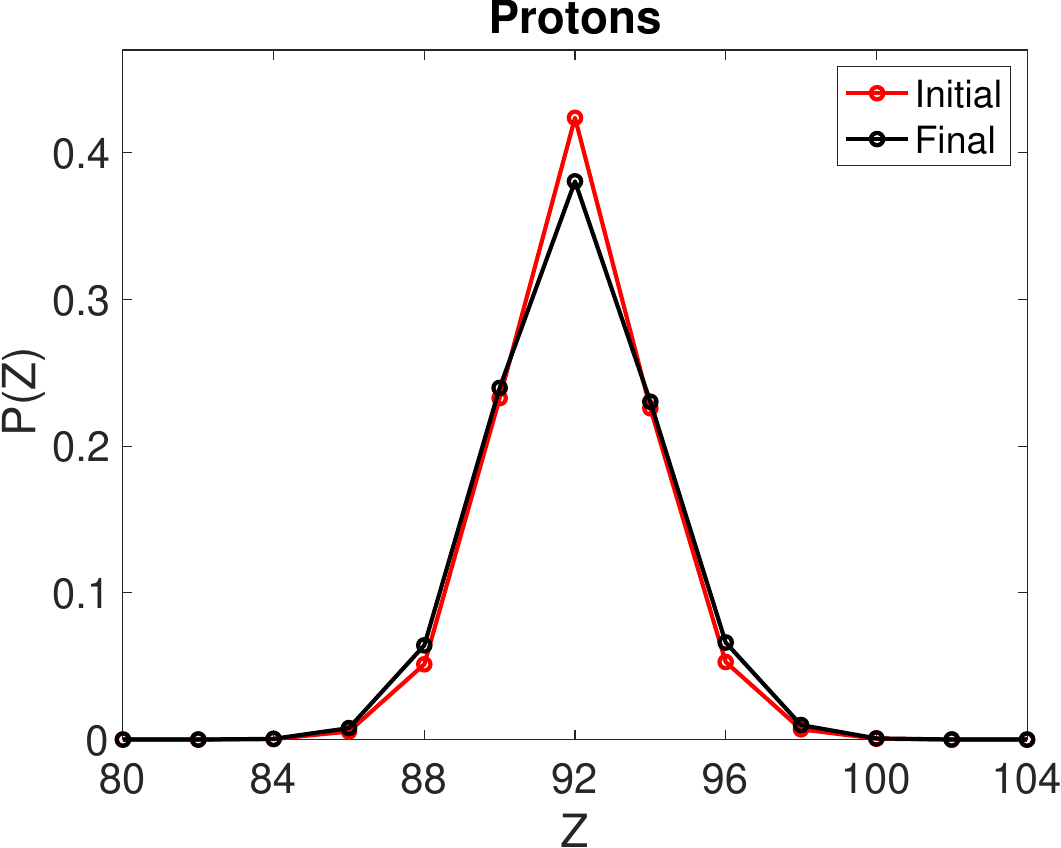}
\includegraphics[width=1.0\columnwidth]{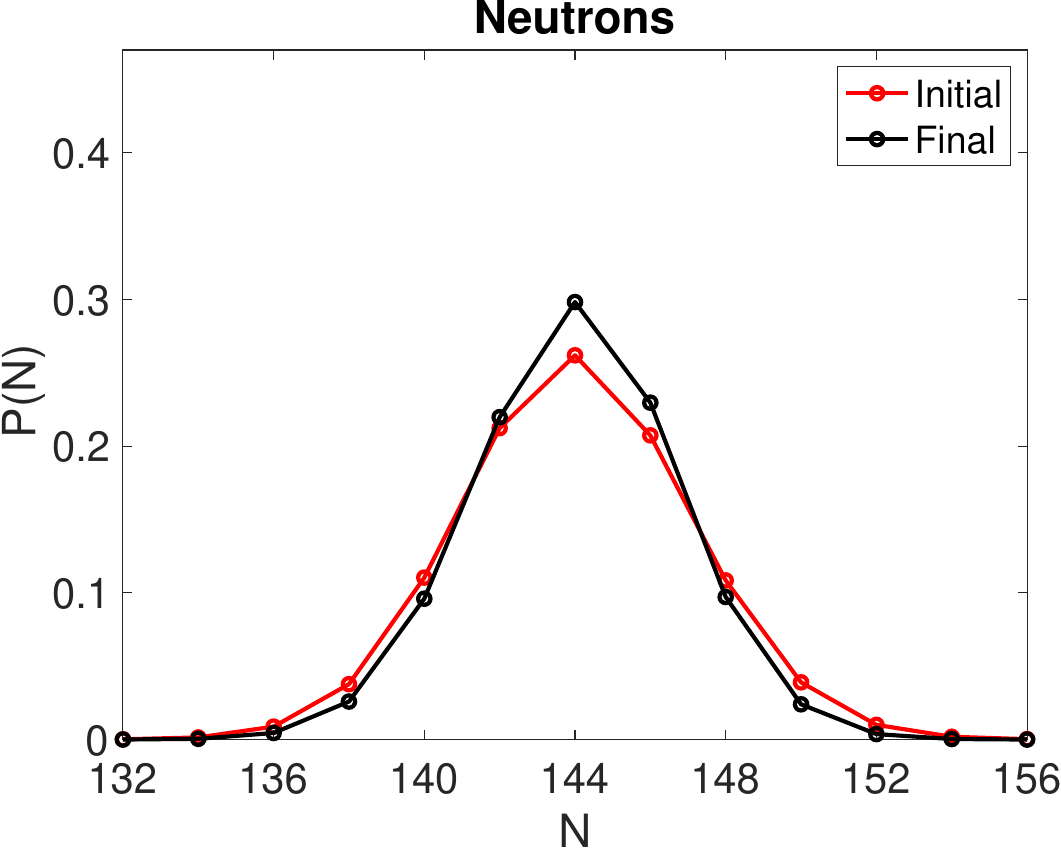}
\caption{ \label{fig:ProjProb} 
In the upper and lower panels we show the probabilities $P(N)$ defined in Eq.~\eqref{equ:N} for protons and neutrons respectively, 
in the initial and final states and used in producing Figs.~\ref{fig:dens1}-\ref{fig:dens3}, without any energy boost, as discussed in the text.}
\end{figure}  
\begin{figure}
\includegraphics[width=1\columnwidth]{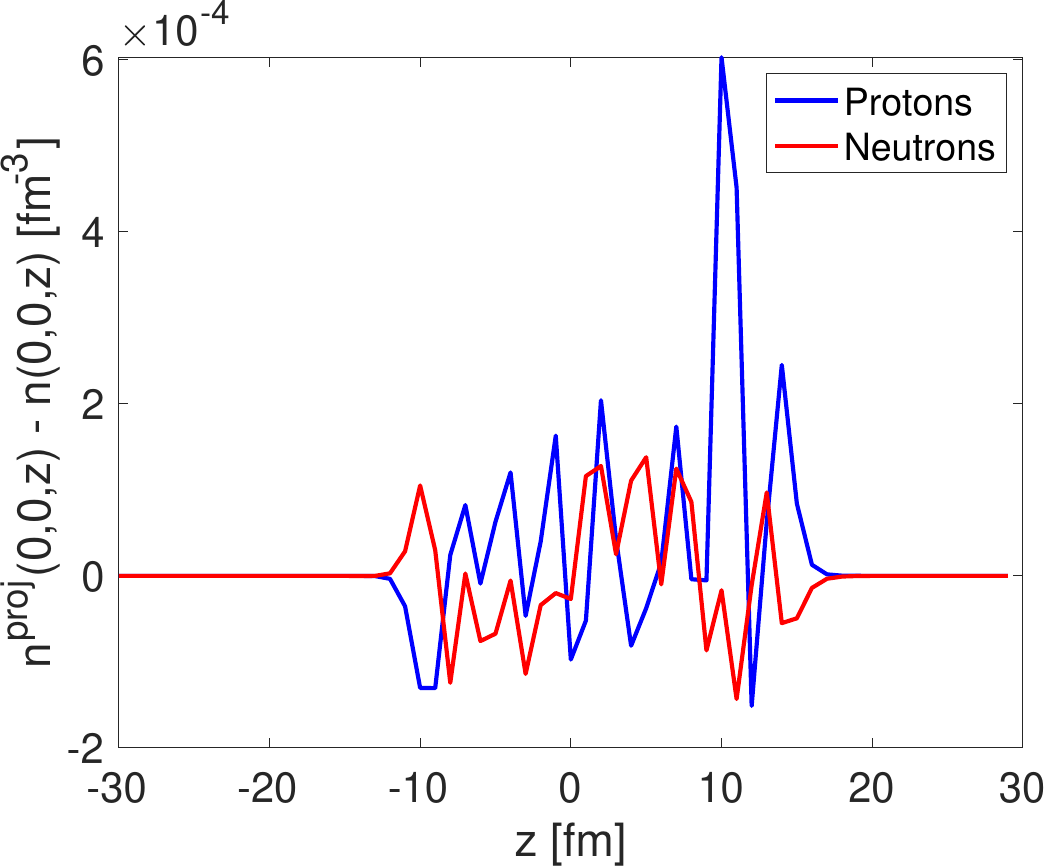}
\includegraphics[width=1\columnwidth]{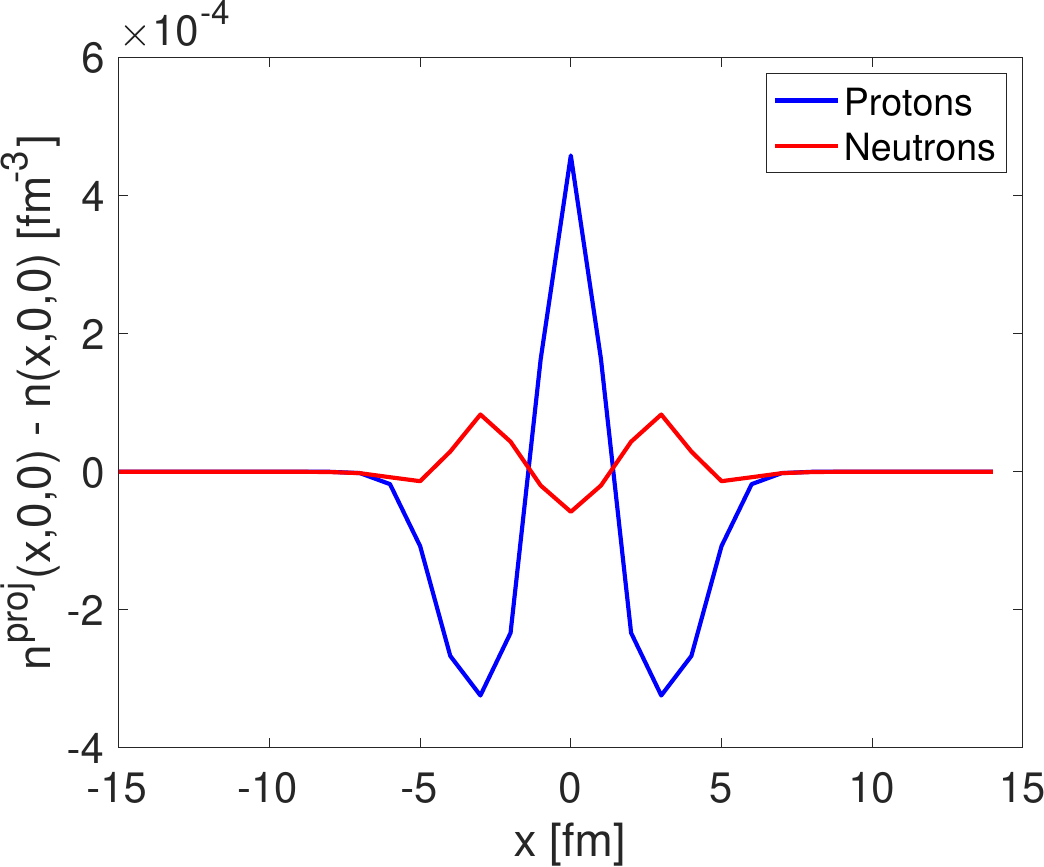}
\caption{ \label{fig:dens1}  
 Differences between unprojected and particle number projected number densities at the top of the outer barrier  for $^{236}$U.  }
\end{figure}  
\begin{figure}
\includegraphics[width=1.0\columnwidth]{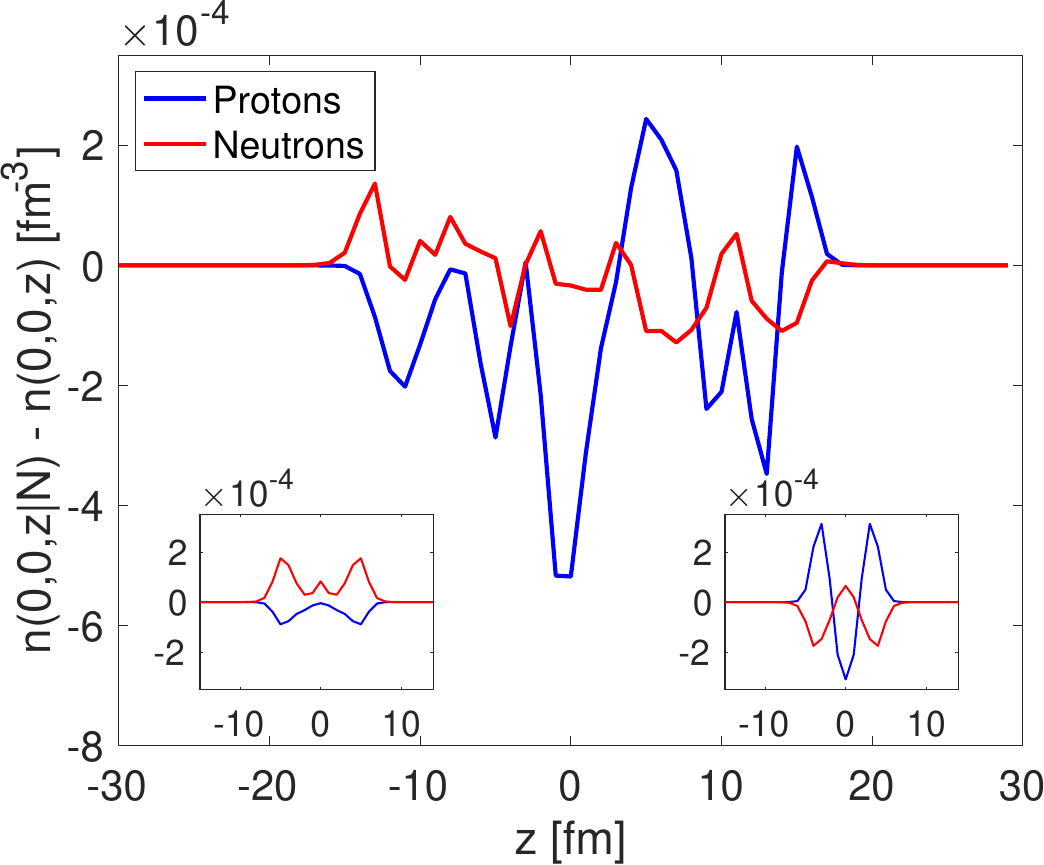}
\caption{ \label{fig:dens2}  
 Differences between unprojected and particle number projected number densities at the scission configuration, 
 for the entire system along the fission direction $z$-axis 
 and  along the $x$-axis and at $y=0$ in the inserts,
 centered at the heavy on the left and 
 at the light on the right fission fragments respectively at scission. In this case and in Figs.~\ref{fig:dens1} and \ref{fig:dens3} 
 the initial state was the same as in Table~\ref{table:initial}, but without a boost of the collective quadruple mode. }
\end{figure}
\begin{figure}
\includegraphics[width=1.0\columnwidth]{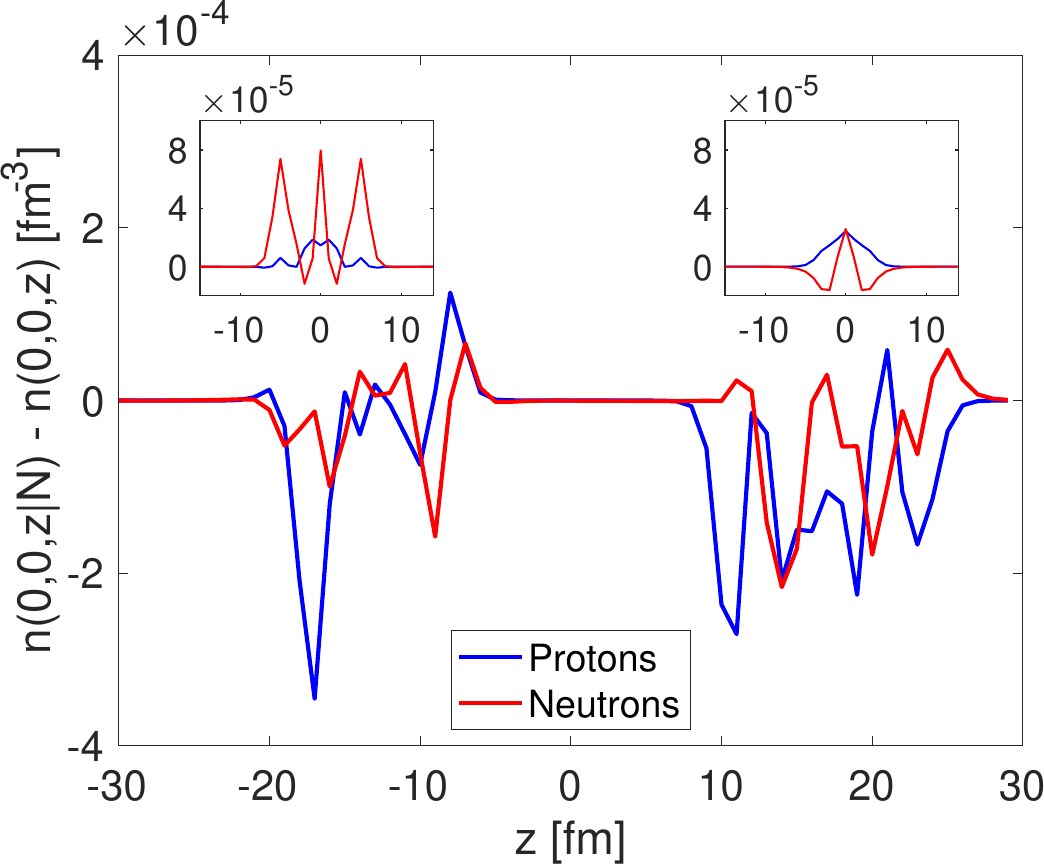}
\caption{ \label{fig:dens3}  
The same as in Fig.~\ref{fig:dens2} but for the fully separated FFs.
 }
\end{figure}

The projected particle number probability distributions $P(N)$, see Eq.~\eqref{equ:N}, for protons and neutrons obtained are shown in Fig.~\ref{fig:ProjProb}, 
are slightly asymmetric with respect to the average proton $Z=92$ and neutron $N=144$ numbers. 
In Figs.~\ref{fig:dens1}-{\ref{fig:dens3} we show the differences between the unprojected and particle-number-projected 
number densities for protons and neutrons in the case of the induced fission of $^{236}$U at the top of the outer barrier, 
at the scission configurations, and for fully separated FFs. 
From the time-dependent full set of QPWFs we
evaluated particle-number projected  densities with only 5\,000 canonical 
quasi-particle wave functions.  The agreement between the particle-number projected and unprojected densities 
at each selected time during the evolution had 
very small relative errors, see also Fig.~\ref{fig:ProjDensProj}.

\subsection{Irreversibility in isolated quantum systems}\label{ssec:VID}

Another relevant aspect for the nuclear dynamics, 
which can be revealed with the help of canonical wave functions, is the irreversible 
time evolution of an isolated nuclear system, before it emits any nucleons or before 
it couples with the electromagnetic fields and emits photons or later on $\beta$-particles 
after coupling with the weak interactions. An isolated excited nucleus has a
vanishing von Neumann or Shannon entropy, which naively would point to the absence of any irreversible 
time-evolution of such a system, which clearly is incorrect.  At the classical level one can evaluate the 
Boltzmann entropy of an excited system, which would clearly characterize the irreversible time evolution 
of the system.  The non-equilibrium evolution of isolated quantum systems can be characterized however 
with the help of the entanglement entropy~\cite{Calabrese:2005,Calabrese:2006,Alba:2017}. The 
entanglement entropy is non-vanishing even in the ground states of interacting systems~\cite{Srednicki:1993}.
There is  however no unique definition of the entanglement entropy, as it is well known~\cite{Nordheim:1928,
Uehling:1933,Srednicki:1993,Klich:2006,Boguslawski:2014,Klich:2006,Amico:2008,
Horodecki:2009,Haque:2009,Eisert:2010,Boguslawski:2014,Gigena:2015,Bengtsson:2017,Johnson:2018,
Robin:2021,Hoppe:2021,Johnson:2022,Bulgac:2022,Bulgac:2022a,Bulgac:2022d,Tichai:2022,Fasano:2022,
Bulgac:2023,Robin:2023,Gu:2023,Pandharipande:1984,Stoitsov:1993,Reinhard:1999,Dobaczewski:1996,Tajima:2004,
Tichai:2019,Fasano:2022,Chen:2022,Hu:2022,Hagen:2022,Kortelainen:2022}. For nuclear systems, and particularly for heavy nuclei,  
which have an enormous number of degrees of freedom only the orbital entropy can be evaluated in the near future
\begin{align} 
S(t) = &- g\sumint_k n_k(t)\ln n_k(t) \nonumber \\
&- g\sumint_k  [1-n_k(t)]\ln[1-n_k(t)],\label{eq:ent}
\end{align}
where $g$ is the spin-isospin degeneracy and the number of single-particle canonical occupation probabilities
is of the order of  ${\cal O}(10^{4\ldots 5})$, determined at each time shown  with a symbol in Fig. \ref{fig:entropy}. The 
entanglement entropy in addition provides an insight on the complexity, or the minimal number of independent 
Slater determinants, required to accurately describe a dynamic process as a function of time~\cite{Bulgac:2023}. 
At different times during the evolution the complexity of the many-body wave function changes, depending on how 
effective is the repopulation of the single-particle states due to the particle-particle interactions, beyond the naive mean field. 
In a simple one Slater determinant time-dependent approximation, known as TDHF, 
the single particle occupation probabilities do not change in time. 

Other authors have considered higher order entropies, 
such as two-body entropies~\cite{Robin:2021,Robin:2023,Gu:2023}, however, only in much smaller 
single-particle spaces than what is needed to simulate dynamics of complex nuclei, such as fission. 
The single-particle occupation probabilities are not well-defined, as their values depend of the basis 
set one uses in order to evaluate them, and a superfluid system is particular example. As it is well known 
for decades now, the canonical wave functions or the natural 
orbitals~\cite{Lowdin:1956,Lowdin:1956a,Coleman:1963,Coleman:1963a} are the smallest possible 
set to represent a many-body wave functions in terms of single-particle orbitals as a sum over $N$-particle 
Slater determinants, as  in particular is needed in shell-model calculations~\cite{Johnson:2018}. For example, 
evaluating the entanglement two-body entropy~\cite{Robin:2021} would require defining the two-body density matrix 
\begin{align}
n_2(\xi,\zeta,\zeta',\xi') =\langle\Phi|\psi^\dagger(\xi')\psi^\dagger(\zeta')\psi(\zeta)\psi(\xi)|\Phi\rangle,
\end{align} 
which in the case of nuclear systems simulated on a spatial lattice, is an object with $(2\times2\times N_xN_yN_z)^4$ 
coordinates,  a quantity too large to fit in any classical supercomputers in the foreseeable future. 
\begin{figure}
\includegraphics[width=1.0\columnwidth]{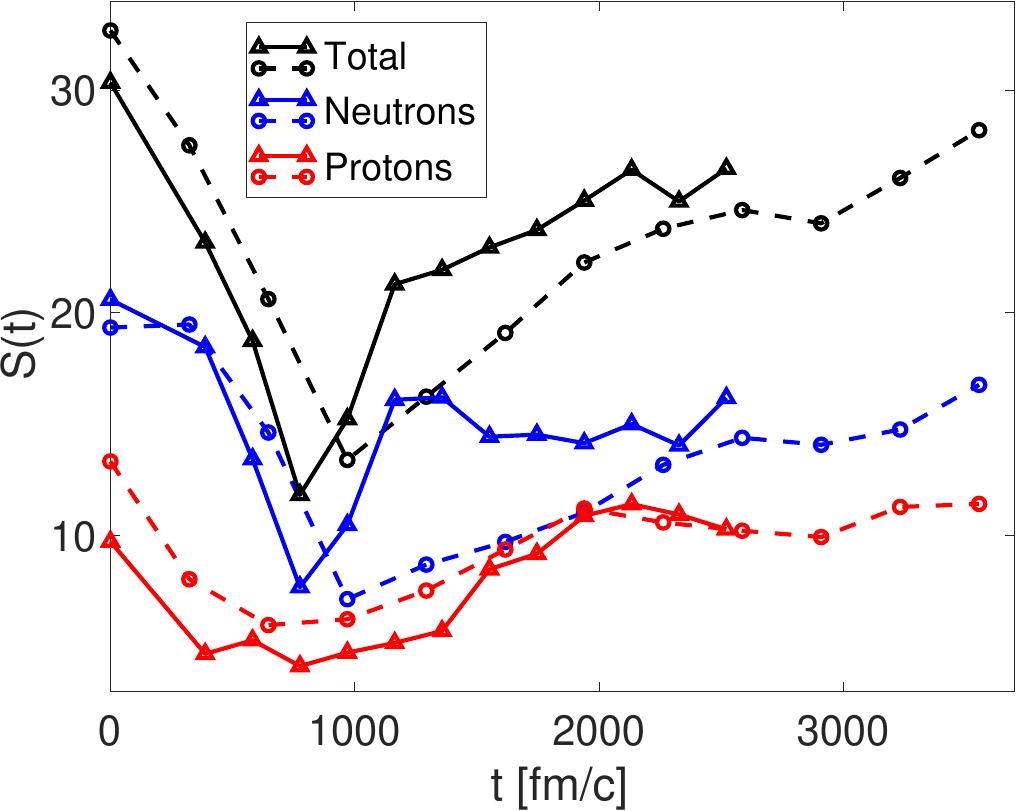}  
\caption{ \label{fig:entropy}  
 The orbital entropy $S(t)$ as a function for time for fission of $^{236}$U in a large simulation box 
 $48^2\times120$ fm$^3$ for two of the simulations 
 reported in Ref.~\cite{Abdurrahman:2023}, 
 for two different set of initial conditions. The initial deformations of the fissioning nuclei were 
 $Q_{20}=159.64$ b  and $Q_{30}=17.80$ b$^{3/2}$ for one trajectory (solid lines) and 
 $Q_{20}=135.25$ b  and $Q_{30}=12.44$ b$^{3/2}$ for the other trajectory (dashed lines), near the top of 
 the outer barrier. The entropy is plotted separately for proton and neutron subsystems 
 and for the entire fissioning system.  For more details concerning these simulations see Ref.~\cite{Abdurrahman:2023}.}
\end{figure}  



\begin{figure}
\includegraphics[width=1.0\columnwidth]{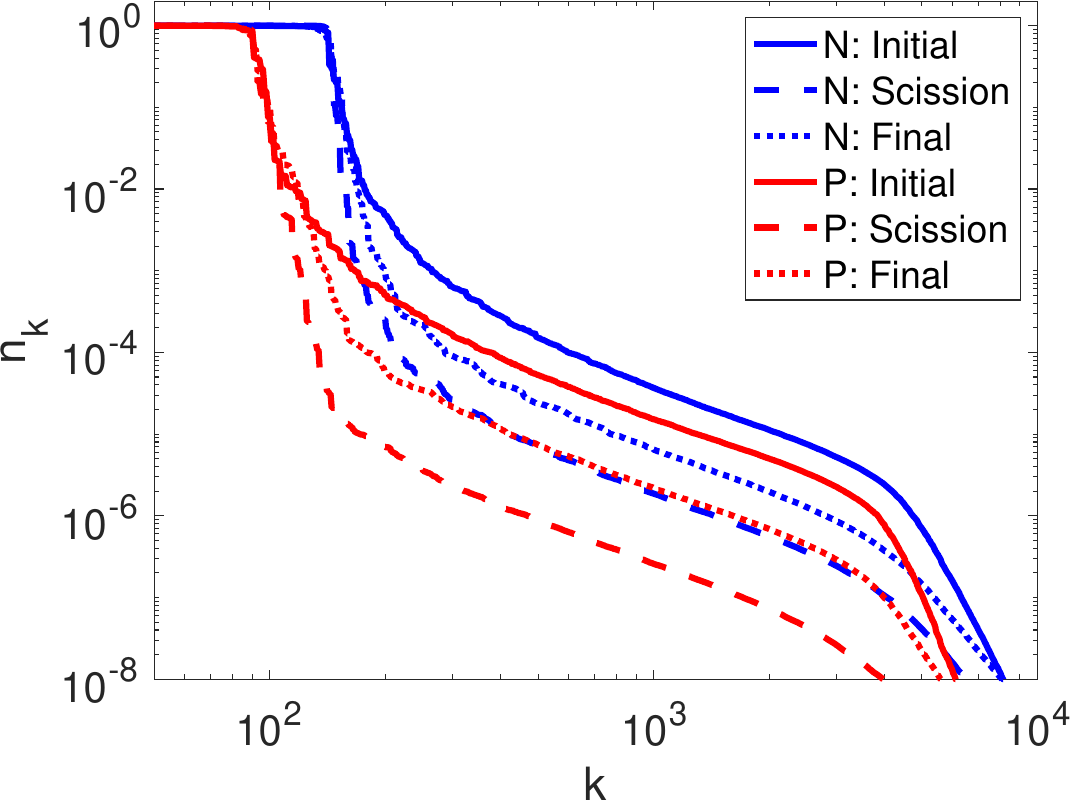}  
\includegraphics[width=1.0\columnwidth]{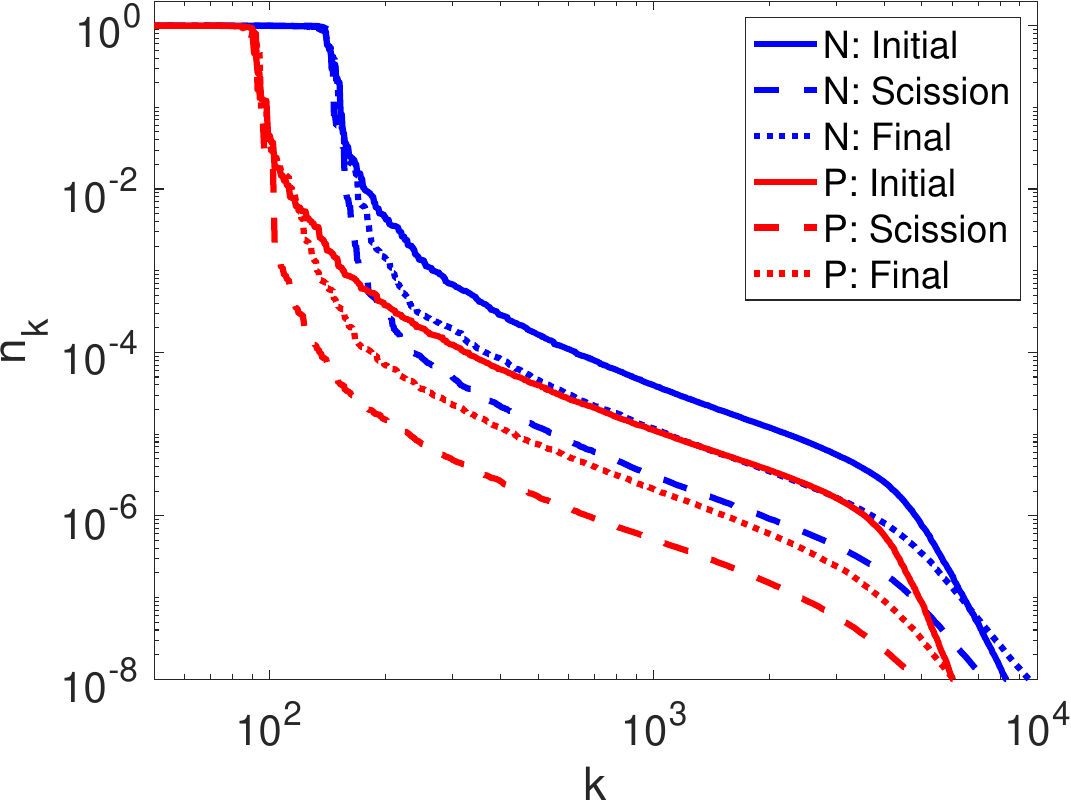}  
\caption{ \label{fig:Occup} 
The time evolution of the canonical occupation probabilities for neutron and proton 
levels at the initial time, scission, and the final time, for the same data presented in Fig.~\ref{fig:entropy}. The upper panel correspond 
the trajectory shown with dashed limes, while the lower panel corresponds to the trajectory shown with solid lines in Fig.~\ref{fig:entropy}. }
\end{figure}  

Here we will illustrate the irreversible fission dynamics in the largest simulation we have performed on a spatial lattice $48^2\times120$, 
which required 4\,609 nodes with 27\,654 GPUs on the Summit supercomputer for about 15 wall-hours 
for a single fission trajectory, in one of the largest (if not the largest) direct numerical simulation ever reported, see Fig.~\ref{fig:entropy}.  Similar results have been reported in Ref.~\cite{Bulgac:2022,Bulgac:2022a,Bulgac:2022d,Bulgac:2023} for both fission and for a $^{238}$U+$^{238}$U collision at 1\,500 MeV in center-of-mass frame, however for much smaller simulation boxes. 
The orbital entanglement entropy is very large initially, as expected in a system with very strong pairing correlations. 
As the nucleus evolves towards scission it heats up, achieving temperatures above 1 MeV~\cite{Bulgac:2016,Bulgac:2019c,Bulgac:2020} 
and the pairing correlations weaken, but they do not disappear, see Refs.~\cite{Bulgac:2016,Bulgac:2019c,Bulgac:2020} and Fig.~\ref{fig:DensEvol}. As \textcite{Magierski:2022} have recently shown, 
even in the high energy collisions of $^{90}$Zr+$^{90}$Zr, due to the excitation of the Higgs pairing mode~\cite{Barankov:2006,Bulgac:2009}, 
pairing correlations, which are absent in the initial nuclei, acquire a large amplitude at very large excitation energies in the proximity 
of the Coulomb barrier of the colliding nuclei  of the compound nucleus formed in this collision.  While approaching the 
scission configuration, the matter exchange between the two halves of the fissioning nucleus slows down and completely stops after scission, 
which in the Fig.~\ref{fig:entropy} is around 1\,000 fm/c. After the two FFs separate, they are highly excited, and in particular the light 
FF is also very highly deformed, and both fragment relax, and the entanglement entropy increases with time, as the single-particle levels 
are repopulated, reaching values 
almost equal to the initial entanglement entropy of the cold strongly correlated nucleus near the top of the outer barrier. 
The temperature of the final FFs is significantly larger, the remaining pairing correlations are weaker, and while one process favors more 
occupation particle redistribution, the other works in the opposite direction. 

In Fig.~\ref{fig:nkdiff2} and subsection~\ref{ssec:VIB} we observed the non-Markovian behavior of the quantity $\sigma_1(t)= \sum_k |n_k(t)-n_k(0)|$, which had a higher essentially linear rate of change before the neck is formed, and a slower, also almost linear rate for larger times. Between the initial state and scission the canonical occupation probability spectrum acquires a somewhat sharper Fermi surface, see Fig. \ref{fig:Occup}, which explains why the orbital entropy decreases. In the final state the canonical occupation probability spectrum moves towards that of the initial state, which again explains why the orbital entanglement entropy increases. These two time-dependent behaviors of the $\sigma_1(t)$ and $S(t)$ are thus correlated and now we see how.    


In the large simulation box we studied the fission of $^{236}$U we thus have clearly identified two regimes. In this very large simulation box the FFs still evolve in time when they reach the box walls and they clearly did not reach thermalization. One might be tempted to isolate each FF separately and follow its evolution in its own center-of-mass. This can appear physically motivated, as at sufficiently large spatial separations one expects that the FFs can hardly influence each other any longer, apart from the relatively weak Coulomb field. It is however not clear how one can formally proceed, since apart from the ${\text v}_k(\xi,t)$ quasiparticle wave functions, which can be very well localized inside a specific FF, the time evolution of a specific FF is also controlled by the 
${\text u}_k(\xi,t)$ component of the quasiparticle wave functions, which are mostly fully delocalized, as we discussed in Section~\ref{sec:III}, and through these components the two FFs can ``communicate'' with each other. This formal aspect of the TDDFT has not been developed yet.

This quantum non-equilibrium entanglement entropy~\cite{Calabrese:2005,Calabrese:2006,Alba:2017} has an unexpected behavior at first sight, 
but this behavior is also observed in the evolution of other much simpler systems of strongly interacting fermions ~\cite{Milburn:1997,
Chuchem:2010,Cohen:2016,Abanin:2019,Sinha:2020,Del-Maestro:2021,Del-Maestro:2022,Thamm:2022}. As a final remark, even though we have illustrated the dynamics of a heavy nucleus with only a few examples, the same features were observed for several hundred fission trajectories and heavy-ion collisions we have performed over the years for various actinides and combination of heavy-ions, the latest still unpublished.

While we have not shown the occupation probability spectrum at scission, qualitatively it looks somewhat similar to the final occupation probability in the final state shown in Fig.~\ref{fig:PartNumb}. As we have discussed above, even if one starts a dynamical simulation with a set of canonical wave functions, they cease to be canonical at the next time step. The canonical occupation probability spectrum has to be determined separately at each time is needed, following the procedure outlined in Section~\ref{sec:II}. In Fig.~\ref{fig:Occup} we show canonical occupation probabilities, needed to evaluate the orbital entropy $S(t)$ at the initial, scission and final time. Unlike the final occupation probability spectrum shown in Fig.~\ref{fig:PartNumb} the canonical occupation probability has a qualitatively similar character at any time. At scission however, both neutron and proton canonical occupation probabilities show noticeably shorter tails, which explains the non-monotonic behavior of the orbital entropy $S(t)$ illustrated in Fig.~\ref{fig:entropy}.

\section{Conclusions}\label{sec:VII}

The use of a reduced set of canonical wave functions/natural orbitals could be a good approximation for treating a variety of static problems, 
when a reduced set of single-particle 
states with non-negligible occupation probabilities above a certain threshold are chosen. Unlike the normal number density, 
the anomalous number density and the kinetic energy density are strictly diverging in the case of local pairing potentials~\cite{Bulgac:1980}, 
since for large energies in 3D the single particle occupation probabilities behave as $n(\varepsilon(p)) \propto 1/\varepsilon^2(p) \propto 1/p^4$, 
and regularization and renormalization are required in order to ensure accurate and reproducible results. The $1/p^4$ behavior of the 
canonical occupation probabilities is cutoff at momenta of the order of $\Lambda_{QCD}$ in nuclear systems or at $\hbar /r_0$, where $r_0$ 
is of the order of the range of the size of the particles.
The canonical wave functions are also very useful when performing particle and/or angular momentum projections~\cite{Scamps:2023a}. 

We have presented compelling arguments for the use of a full set of quasi-particle wave functions in time-dependent density 
functional theory simulations. The use of a smaller set of quasi-particle wave functions, or approximations such as TDBCS, lead to 
incorrect results or even fail to fission entirely. In a correct implementation of the dynamics, quasi-particle levels with even very small occupation probability, including 
those completely negligible at an initial time, quite often are populated at a later time
to such a level that the final results could be qualitatively different between approximate and exact results.
One should remember that the TDBCS approximation, which is a further approximation of the full TDHFB, is still 
using a reduced set of initial canonical quasi-particle wave functions, an approximation which leads to errors, as we have shown here.   This approach 
quite widely used by a number of practitioners, which apart from violating the 
continuity equation, and thus failing to correctly describe the nuclear shape evolution, can lead to results quite different from the 
full TDDFT framework.  In systems with strong pairing, such as nuclear systems and cold atom systems, the $u-$ and $v-$components 
of the Bogoliubov quasi-particle wave functions, where the $u$-component often lies in the continuum, or the exact time-reverse canonical 
orbitals $\phi_k(\xi)$ and $\phi_{\overline k}(\xi)$ have very different spatial profiles and cannot be treated in the BCS approximation.  

We have demonstrated that using static self-consistent solutions of the DFT including pairing correlations with local pairing potentials lead 
different results, depending on whether one uses a BCS or a full HFB implementation of pairing correlations, the final solution depends quite 
strongly on the level of spatial resolution adopted, either by using a spatial lattice or a set of appropriately rescaled harmonic oscillator wave functions, 
as in the very popular code HFBTHO~\cite{Navarro:2017,Marevic:2022}. This is an important aspect of defining 
various nuclear energy density functionals, since because of the
inherent divergence of the anomalous density~\cite{Bulgac:1980}, the self-consistent equations have to be regularized and 
renormalized~\cite{Bulgac:2002,Bulgac:2002a} in each specific numerical implementation, in total analogy with running coupling constants in 
quantum field theory, and it is not enough to specify the values of the coupling constants alone, but also the equivalent spatial resolution used 
in order to obtain nuclear masses, charge radii and other nuclear properties. This aspect becomes even more important 
in time-dependent simulations, since depending on the level of spatial resolution, the available phase space varies significantly and 
so does the dynamical evolution and the instantaneous single-particle occupation probabilities. Ignoring these aspects, particularly in 
simulation at relatively low spatial resolutions leads to vastly different properties of the final states, and thus the confrontation of 
the theory with experimental data becomes questionable.

Finally, we have shown how the use of time-dependent canonical occupation probabilities 
allows the determination of the orbital entanglement 
entropy, which provides insight into the irreversible dynamics of isolated quantum systems. This, in particular in the case of fission dynamics, also gives information 
about the time-dependence of the complexity of the many-body wave function of a strongly interacting system.  The presence of pairing correlations within the TDDFT extension is tantamount to the presence of a quantum collision integral in the evolution equations~\cite{Bulgac:2022}, which leads to an obviously non-Markovian behavior, unexpected in the presence of strong dissipation in a traditional \textcite{Nordheim:1928} and \textcite{Uehling:1933} formulation of the quantum kinetic theory. The extension of the TDDFT framework to superfluid systems has similarities with the Baym and Kadanoff  extension framework~\cite{Baym:1961,Baym:1962}, see also the independent work of \textcite{Keldysh:1965}. These extensions of the non-equilibrium dynamics are however much more complex as they rely on very complex memory and non-local kernels, and their application to such a complex phenomenon as nuclear fission, would be numerically impossible in the foreseeable future. 
Unlike the von Neumann or 
Shannon entropy, which vanishes for an isolated quantum system and thus fails to describe the irreversible dynamics and 
the expected thermalization of the excited nuclei, the  orbital entanglement entropy is likely the most useful characterization of the 
quantum dynamics of an isolated nucleus, which can be 
evaluated for rather complex time-dependent many-body systems.

 \vspace{0.5cm}

{\bf Acknowledgements} \\

We have benefited at various times from discussions on issues discussed here with K. Godbey, G. Scamps, and A. Makowski. 
The funding for AB from the Office of Science, Grant No. DE-FG02-97ER41014  
and also the partial support provided by NNSA cooperative Agreement DE-NA0003841 is greatly appreciated. 
MK was supported by  NNSA cooperative Agreement DE-NA0003841.
This work was carried out under the auspices of the National Nuclear Security Administration of the U.S. Department of Energy at Los Alamos National Laboratory under Contract No. 89233218CNA000001, and used resources of the Oak Ridge Leadership Computing Facility, which is a U.S. DOE Office of
Science User Facility supported under Contract No. DE-AC05-00OR22725. 
I.A. and I.S. gratefully acknowledges partial support and computational resources
provided by the Advanced Simulation and Computing (ASC) Program.

%
\providecommand{\selectlanguage}[1]{}
\renewcommand{\selectlanguage}[1]{}

\bibliography{local_fission1}

\end{document}